\documentclass[twocolumn, astrosymb, times, twocolappendix]{aastex631}

\usepackage{amsmath}
\usepackage{mathptmx}
\usepackage{mdwlist}
\usepackage{ulem}
\usepackage{url}
\usepackage{xspace}
\usepackage{xcolor}

\let\tablenum\relax
\usepackage{siunitx}

\newcommand{\acro}[1]{\textsc{#1}\xspace}
\newcommand{\things}{\acro{things}}
\newcommand{\heracles}{\acro{heracles}}
\newcommand{\bimasong}{\acro{bima~song}}
\newcommand{\sings}{\acro{sings}}
\newcommand{\sfourg}{\acro{s\textsuperscript{4}g}}

\newcommand{\qtyerr}[2]{#1\pm #2}
\newcommand{\qtyerrs}[3]{#1^{+#2}_{-#3}}
\newcommand{\ngc}[1]{\object[NGC#1]{NGC\,#1}\xspace}
\newcommand{\messier}[1]{\object[M#1]{M\,#1}\xspace}

\DeclareSIUnit{\solarmass}{M_\sun}
\DeclareSIUnit{\solarluminosity}{L_\sun}
\DeclareSIUnit{\magnitude}{mag}
\DeclareSIUnit{\parsec}{pc}
\DeclareSIUnit{\pixel}{pixel}
\DeclareSIUnit{\jansky}{Jy}
\DeclareSIUnit{\beam}{beam}
\DeclareSIUnit{\yr}{yr}
\DeclareSIUnit{\dex}{dex}
\DeclareSIUnit{\arsc}{arcsec}

\definecolor{orangered}{RGB}{255,69,0}

\received{...}
\revised{...}
\accepted{...}

\submitjournal{ApJ}

\begin{document}

\title{Connection between Non-Axisymmetric Structures and Neutral Gas Distribution in Disk Galaxies}

\author{Ze-Zhong Liang}
\affiliation{Department of Astronomy, School of Physics, Peking University, Beijing 100871, People's Republic of China}

\author[0000-0002-6593-8820]{Jing Wang}
\affiliation{Kavli Institute for Astronomy and Astrophysics, Peking University, Beijing 100871, People's Republic of China}

\author[0000-0003-1015-5367]{Hua Gao}
\affiliation{Institute for Astronomy, University of Hawaii, 2680 Woodlawn Drive, Honolulu, HI 96822, USA}

\author[0000-0001-6947-5846]{Luis C. Ho}
\affil{Kavli Institute for Astronomy and Astrophysics, Peking University, Beijing 100871, People's Republic of China}
\affil{Department of Astronomy, School of Physics, Peking University, Beijing 100871, People's Republic of China}

\author[0000-0001-6079-1332]{E. Athanassoula}
\affiliation{Aix Marseille Univ, CNRS, CNES, LAM, Marseille, France}

\correspondingauthor{Jing Wang}
\email{jwang\_astro@pku.edu.cn}



\begin{abstract}

Non-axisymmetric structures, such as bars and spiral arms, are known to concentrate molecular gas and star formation in galaxy centers, actively building up the pseudo-bulges. However, a direct link between the neutral (i.e., molecular and atomic) gas distribution and the exerted torque forces over a broader radial range of galactic disks still remains to be explored. In the present work, we investigate this link by carefully evaluating the torque force field using the \qty{3.6}{\micro \meter} images for \qty{17}{} The \ion{H}{1} Nearby Galaxy Survey (\things) galaxies, and measuring neutral gas distribution on resolved atomic and molecular line maps. We find that galaxies with stronger torque forces show a more concentrated neutral gas distribution over the disk-scale, defined as half the isophotal radius at \qty{25.5}{\magnitude \per \square \arsc}. The correlation holds regardless of whether the neutral gas fraction, or the effective stellar mass surface density is controlled for. In addition, \qty{}{\kilo \parsec}-scale neutral gas over-densities tend to be located close to the local maxima of torque forces. Most of these correlations involving the torque forces are comparatively stronger than those using the traditional Fourier amplitudes to quantify the non-axisymmetric structures. These results are consistent with the scenario that non-axisymmetric structures exert torque forces, and trigger dissipative processes to transport gas inward, not only to build the pseudo-bulges, but also fuel the inner disk growth. In this regard, non-axisymmetric structures inducing stronger torque forces appear to be more efficient in these processes.

\end{abstract}



\section{Introduction} \label{sec:intro}

As redshift decreases, galaxy evolution becomes much slower \cite[]{2012ApJ...757...60K}, and non-axisymmetric structures, such as bars and spiral arms, are expected to play a significant role in this secular phase of galaxy evolution \cite[]{2004ARA&A..42..603K, 2016MNRAS.462.3430K, 2022ARA&A..60...73S}.

Bars, for example, have long been considered to be a major driver of galaxy evolution. Observations have demonstrated that bars are ubiquitous in the local Universe \cite[e.g.,][]{2008ApJ...675.1141S, 2018MNRAS.474.5372E, 2018ApJ...862..100G}. Moving to higher redshifts, recent \textit{James-Webb Space Telescope} (\acro{jwst}) observations have found bars up to $z\approx 3$, suggesting that the bar-related processes might have already started at these early epochs \cite[e.g.,][]{2023ApJ...945L..10G, 2023ApJ...958L..26H, 2024MNRAS.530.1984L}. Theories have predicted that the interaction between bars and gas would significantly impact the galaxy evolution \cite[]{2003MNRAS.341.1179A, 2006ApJ...645..209D, 2010ApJ...721.1878S}. Forces exerted by bars could induce shock processes that effectively transport the gas inward \cite[e.g.,][]{1992MNRAS.259..345A, 2012ApJ...758...14K, 2015MNRAS.449.2421S}. In simulations, this inflow could at an inflow rate of $\sim$ \qtyrange[range-phrase=--, range-units=bracket]{0.01}{100}{\solarmass \per \yr}, depending on bar strength and pattern speed \cite[]{1992MNRAS.259..345A, 2004ApJ...600..595R, 2012ApJ...758...14K}. This gas inflow might fuel central star formation, and a central mass concentration of gas and stars formed out of the gas with mass up to $\sim \qty{e+10}{\solarmass}$ might form in the form of disk-like pseudo-bulges \cite[e.g.,][]{2005MNRAS.358.1477A, 2013MNRAS.429.1949A, 2014MNRAS.445.3352C}. Note that \cite{2020ApJS..247...20G} have found neither more abundant nor more prominent pseudo-bulges in barred galaxies compared to the non-barred counterparts, likely because of the self-limit nature of this fuelling processes, though one needs to keep in mind the uncertainties in classifications of bulges due to disagreement in the adopted criteria \cite[]{2009MNRAS.393.1531G, 2017A&A...604A..30N, 2022ApJS..262...54G}. The gas might also stalls due to dynamical resonances, shocks, increased centrifugal force or reduced shear \cite[e.g.,][]{2000A&A...362..465R, 2003ApJ...582..723R,  2012ApJ...758...14K, 2015MNRAS.453..739K}, leading to local accumulation of gas in the form of rings and other gas morphological features \cite[]{1996FCPh...17...95B, 2012ApJ...747...60K}.

Previous observations focusing on barred galaxies have indeed found kinematic evidence supporting the inward motion of gas \cite[]{2016MNRAS.457.2642S, 2021ApJ...923..220D}. Both \textit{molecular} gas distribution \cite[e.g.,][]{1999ApJ...525..691S, 2005ApJ...632..217S, 2019MNRAS.484.5192C} and star formation rate at timescales of up to $\sim \qty{2}{\giga \yr}$ \cite[e.g.,][]{2011MNRAS.416.2182E, 2012MNRAS.423.3486W, 2020ApJ...893...19W, 2017ApJ...838..105L} have been observed to be elevated in barred galaxy centre, likely explaining the origin of central disk-like components \cite[e.g.,][]{2014A&A...572A..25M, 2015MNRAS.446.4039E, 2020A&A...643A..14G}. While most previous works focussed on the concentration of molecular gas and star formation, the molecular gas has a short depletion time ($\lesssim \qty{1}{\giga \yr}$), particularly in the galactic central regions \cite[]{2008AJ....136.2782L}. The molecular gas needs to be replenished by the atomic gas that is not only of similar masses as the molecular gas within the star-forming disks \cite[]{2020ApJ...890...63W}, but is also \qtyrange[range-phrase=--, range-units=bracket]{1.5}{4}{} times more extended than the star-forming disks \cite[]{2002A&A...390..829S, 2020ApJ...890...63W}. The atomic gas should be susceptible to the same inflow driven mechanisms as the molecular gas, and its inflow should benefit the growth of not only bulge masses, but also disk mass on its way. Interestingly, both simulations and observations support that bars would simultaneously lead to a central hole in \textit{atomic} gas (\ion{H}{1}) distribution \cite[]{2013MNRAS.429.1949A, 2020MNRAS.492.4697N}. Therefore, the net effect of bars in funnelling the \textit{neutral} (atomic \textit{and} molecular) gas remains elusive. Observations have also revealed rich gas morphological features in barred galaxies \cite[]{2021ApJ...913..113M, 2023A&A...676A.113S}, yet their link with the non-axisymmetric structures and their association with disk growth remains to be unveiled.

Spiral arms, being another generic morphological feature in the local Universe, could also play a role in driving galaxy secular evolution \cite[]{2022ARA&A..60...73S}. Spiral arms are also expected to induce shock processes, and funnel the gas inward \cite[]{2014MNRAS.440..208K}. \cite{2014MNRAS.440..208K} have estimated the average inflow rate induced by spiral potential to be $\sim$ \qtyrange[range-phrase=--, range-units=bracket]{0.05}{0.3}{\solarmass \per \yr}, also depending on the spiral strength and the pattern speed. Observations have demonstrated that the force of spiral arms might compress the gas dissipatively \cite[]{2021ApJ...913..113M} and could transport the gas inward \cite[]{2005A&A...441.1011G, 2009ApJ...692.1623H, 2016A&A...588A..33Q}. More active galaxy star formation, more active central star formation rate, and higher fraction of pseudo-bulges have also been observed in galaxies with stronger spiral arms compared to the weaker counterparts \cite[]{2021ApJ...917...88Y, 2022A&A...661A..98Y, 2022A&A...661A..98Y}.

In a sense, various non-axisymmetric structures (bars, spiral arms, also ovals etc.) share a uniformity in terms of inducing the gas inflow to affect galaxy evolution \cite[]{2004ARA&A..42..603K}, with the efficiency presumably dependent upon the \textit{strength}, i.e., the degree they are able to perturb the regular motion of disk neutral gas. While this uniformity has already been tested observationally using, e.g., stellar mass surface density as a strength proxy \cite[]{2022A&A...661A..98Y}, torque force offers comparatively a more direct characterization \cite[e.g.,][]{1981A&A....96..164C, 1992MNRAS.259..345A, 1994ApJ...437..162Q, 2002MNRAS.337.1118L, 2002MNRAS.330...35A, 2016AAA...587A.160D}. This is because that the latter not only offers a more physical measure of the effects due to non-axisymmetric structure but could also better accounts for (through a normalization by the centrifugal force) the stabilizing effects of the symmetric part of disk, the bulge \cite[]{2004MNRAS.355.1251L} and the dark matter halo \cite[]{2016AAA...587A.160D}, which counters the action of non-axisymmetric structures. Yet, direct correlations between torque force and the \textit{neutral} gas distribution almost remains elusive \cite[but see][]{2007PASJ...59..117K}. Furthermore, it is conjectured that different non-axisymmetric structures could in principle work together or in relay to transport the gas across the disk \cite[]{2010ApJ...722..112M, 2013ApJ...769..100S, 2020ApJ...893...19W}. On this regard, it is possible that \textit{radially average strength} of the non-axisymmetric structures, rather than the maximal force corresponding to the strongest component, that is more important in determining the effectiveness of global gas transport processes, yet this still remains to be tested.

The present work would therefore aim to revisit the link between non-axisymmetric structures characterized by the force they exert, and the distribution of the \textit{neutral} gas, i.e., the atomic plus the molecular gas. Both atomic and molecular data with resolution that is able to resolve the stellar disk is therefore necessary for undertaking this investigation. To this aim, we refer to the \ion{H}{1} Nearby Galaxy Survey (\things{}\@; \citealp{2008AJ....136.2563W}) sample, for which uniform measurements of their \ion{H}{1} and molecular gas content, as well as the stellar mass distribution from the \qty{3.6}{\micro \meter} mid-infrared images, are available \cite[]{2003PASP..115..928K, 2009AJ....137.4670L, 2010PASP..122.1397S}. An inevitable cost for this is naturally a small sample size (\qty{17}{}), which calls for careful statistical analysis as well as meticulous uncertainty estimation. We note that the present work could be expanded in the future, either with a uniform reduction of the archival, high-resolution \ion{H}{1} imaging data, or when more data from \acro{ska} pathfinder surveys, including \acro{mongoose} \cite[]{2024A&A...688A.109D} and \acro{wallaby} \cite[]{2020Ap&SS.365..118K}, are available.

The present work would be organized as follows. In Section~\ref{sec:data}, we briefly describe the sample and the data. In Section~\ref{sec:method} we explain how we evaluate the torque force and quantify the neutral gas distribution. In Section~\ref{sec:result}, we present our results, which are further discussed in Section~\ref{sec:discussion} and summarized in Section~\ref{sec:summary}.

\section{Sample and Data} \label{sec:data}

\subsection{Sample} \label{subsec:sample}


\begin{deluxetable}{ccccccc}
    \tabletypesize{\footnotesize}
    \tablewidth{0pt}
    \tablecaption{Sample Galaxies}
    \label{table:sample}
    \tablehead{
        \colhead{Index}
		& \colhead{Object}
        & \colhead{$\alpha$}
        & \colhead{$\delta$}
		& \colhead{$i$}
		& \colhead{$\phi _{\mathrm{PA}}$}
		& \colhead{Note} \\
        \colhead{}
		& \colhead{}
        & \colhead{(\qty{}{hms})}
        & \colhead{(\qty{}{\degree \arcmin \arcsec})}
		& \colhead{(\qty{}{\degree})}
		& \colhead{(\qty{}{\degree})}
		& \colhead{}
	}
    \colnumbers
    \startdata
    1 & \ngc{628} & 1 36 41.7 & $+$15 47 01.1 & $\qtyerr{28.4}{1.7}$ & $\qtyerr{171}{36}$ & * \\
    2 & \ngc{925} & 2 27 16.5 & $+$33 34 43.5 & $\qtyerr{63.3}{0.4}$ & $\qtyerr{16}{2}$ &  \\
    3 & \ngc{2403} & 7 36 51.1 & $+$65 36 02.9 & $\qtyerr{62.9}{1.1}$ & $\qtyerr{34}{2}$ &  \\
    4 & \ngc{2903} & 9 32 10.1 & $+$21 30 04.3 & $\qtyerr{65.4}{2.7}$ & $\qtyerr{114}{4}$ &  \\
    5 & \ngc{2976} & 9 47 15.3 & $+$67 55 00.0 & $\qtyerr{64.8}{1.8}$ & $\qtyerr{64}{12}$ &  \\
    6 & \ngc{3031} & 9 55 33.1 & $+$69 03 54.7 & $\qtyerr{58.7}{2.6}$ & $\qtyerr{60}{2}$ &  \\
    7 & \ngc{3184} & 10 18 16.9 & $+$41 25 27.6 & $\qtyerr{31.8}{2.9}$ & $\qtyerr{56}{24}$ & * \\
    8 & \ngc{3351} & 10 43 57.8 & $+$11 42 13.2 & $\qtyerr{45.6}{2.4}$ & $\qtyerr{102}{4}$ & * \\
    9 & \ngc{3621} & 11 18 16.5 & $-$32 48 50.9 & $\qtyerr{64.7}{1.0}$ & $\qtyerr{75}{3}$ & $\ddagger$ \\
    10 & \ngc{3627} & 11 20 15.0 & $+$12 59 29.6 & $\qtyerr{61.1}{4.6}$ & $\qtyerr{82}{3}$ &  \\
    11 & \ngc{4736} & 12 50 53.0 & $+$41 07 13.2 & $\qtyerr{37.6}{1.2}$ & $\qtyerr{35}{5}$ & $\dagger$ \\
    12 & \ngc{5055} & 13 15 49.2 & $+$42 01 45.3 & $\qtyerr{64.8}{1.7}$ & $\qtyerr{10}{2}$ & $\dagger$ \\
    13 & \ngc{5194} & 13 29 52.7 & $+$47 11 42.4 & $\qtyerr{42.3}{4.7}$ & $\qtyerr{131}{5}$ & * \\
    14 & \ngc{5236} & 13 37 00.9 & $-$29 51 55.9 & $\qtyerr{19.9}{0.6}$ & $\qtyerr{97}{8}$ & *, $\ddagger$ \\
    15 & \ngc{5457} & 14 03 12.5 & $+$54 20 55.9 & $\qtyerr{29.5}{4.6}$ & $\qtyerr{12}{41}$ & * \\
    16 & \ngc{6946} & 20 34 52.2 & $+$60 09 14.4 & $\qtyerr{32.3}{1.9}$ & $\qtyerr{152}{3}$ &  \\
    17 & \ngc{7793} & 23 57 49.7 & $-$32 35 27.9 & $\qtyerr{52.7}{1.0}$ & $\qtyerr{10}{3}$ & $\dagger$, $\ddagger$ \\
    \enddata
    \tablecomments{Summary of the sample galaxy properties. (1) Index. (2) Object designation. (3)--(4) Central coordinate. (5)--(6) Disk inclination $i$ and position angle $\phi _{\mathrm{PA}}$. (7) Note.\\
    * Central coordinate from \cite{2008AJ....136.2720T} and \ion{H}{1} kinematic model from \cite{2008AJ....136.2648D} are not availabe. We determine central coordinate and disk orientation parameters using \qty{3.6}{\micro \meter} images.\\
    $\dagger$ CO imaging data from \cite{2003ApJS..145..259H} and \cite{2009AJ....137.4670L} are not available. We estimate molecular gas distribution by inverting the sub-\qty{}{\kilo \parsec} Schmidt law (Section~\ref{subsec:molecular}).\\
    $\ddagger$ Strong disk warp is identified from the \ion{H}{1} kinematic model of \cite{2008AJ....136.2648D}.
    }
\end{deluxetable}


As mentioned in the previous Section, we select our sample galaxies from the \things sample, which consists of 34 nearby ($D\lesssim \qty{15}{\mega \parsec}$), late-type (Hubble type $T>0$) galaxies that cover a wide range of galactic properties \cite[]{2008AJ....136.2563W}. We further select disk galaxy sample that satisfy (\romannumeral1) $T< \qty{9}{}$, and (\romannumeral2) inclination $i\lesssim \qty{65}{\degree}$, according to table~1 of \cite{2008AJ....136.2563W}. The latter criterion ensures that de-projection would not introduce significant uncertainties in characterizing the disk structures \cite[]{2014ApJ...791...11Z}.

In addition, we consider \ngc{4826} not an ideal case to study the secular evolution driven by non-axisymmetric structures, since \ngc{4826} has been observed to possess counter-rotating gaseous disks, which should signify a transient state due to external perturbation \cite[]{1992Natur.360..442B, 1994ApJ...420..558B}. Our sample therefore consists of \qty{17}{} disk galaxies as listed in Table~\ref{table:sample}.

\subsection{\qty{3.6}{\micro \meter} Imaging Data} \label{subsec:midinfrared}

We mainly use the \qty{3.6}{\micro \meter} IRAC 1 (\qty{3.6}{\micro \meter}) imaging data from the \textit{Spitzer} Infrared Nearby Galaxies Survey \cite[\sings;][]{2003PASP..115..928K, https://doi.org/10.26131/irsa424}\footnote{\url{https://www.ipac.caltech.edu/doi/irsa/10.26131/IRSA424}} to trace the stellar mass distribution, wherever available. The \sings IRAC 1 imaging data feature a point spread function (PSF) full width half maximum (FWHM) of $\gamma _\mathrm{PSF}\approx \qty{1.66}{\arcsecond}$ or typically $\sim \qty{70}{\parsec}$,\footnote{IRAC PSF properties for the \textit{Spitzer} cryogenic mission is quoted. See section~2.2.2 of \cite{2021..............I}.} and reach a depth typically of $\sim \qty{0.003}{\mega \jansky \per \steradian}$ or $\sim \qty{0.7}{\solarmass \per \square \parsec}$. For \ngc{2903}, \ngc{5236} and \ngc{5457}, we use the IRAC 1 imaging data from the \textit{Spitzer} Survey of Stellar Structure in Galaxies \cite[\sfourg;][]{2010PASP..122.1397S, 2013ApJ...771...59M, 2015ApJS..219....5Q, 2022A&A...660A..69W, https://doi.org/10.26131/irsa425}\footnote{\url{https://www.ipac.caltech.edu/doi/irsa/10.26131/IRSA425}}, which feature a PSF FWHM of $\gamma_{\mathrm{PSF}} \approx$ \qtyrange[range-phrase=--, range-units=bracket]{1.7}{2.1}{\arcsecond} \cite[]{2010PASP..122.1397S, 2015APJS..219....4S}\footnote{PSF properties for the \textit{Spitzer} post-cryogenic mission is quoted, as bulk of the \sfourg imaging data has been obtained over this peroid.} and, in these three cases, similar physical resolutions and depths with the \sings IRAC 1 images

Masks for unrelated sources, and sigma maps, i.e., pixel-wise estimation of flux uncertainty, of \sfourg imaging data are directly adopted from \cite{2013ApJ...771...59M}, since they have already been optimized for the data. For \sings images, we generate masks and sigma maps following the procedures outlined in \cite{2009ApJ...703.1569M} and \cite{2015APJS..219....4S}, respectively\footnote{Note that for \sings images, the background subtracted by the \sings pipeline should also be added back in equation~8 of \cite{2015APJS..219....4S}.}. To strictly mask out unrelated sources, we also make use of \textsc{ProFound} \cite[]{2018ascl.soft04006R, 2018MNRAS.476.3137R}, which is capable of automatically dilating the mask to contain most flux of the unrelated sources, to generate a second mask and combine it with the original one. Finally, the \qty{3.6}{\micro \meter} images is converted to the stellar mass surface density following \cite[]{2013ApJ...771...59M}
\begin{equation} \label{eq:stellar_mass_surface_density}
    \frac{\Sigma _{\star}}{\qty{}{\solarmass \per \square \parsec}}=\qty{3.7e+2}{}F_{\mathrm{corr}}\cos i\left( \frac{I_{\qty{3.6}{\micro \meter}}}{\qty{}{\mega \jansky \per \steradian}} \right) ,
\end{equation}
assuming a \qty{3.6}{\micro \meter} mass-to-light ratio $\Upsilon_\mathrm{\qty{3.6}{\micro \meter}}=0.53$ \cite[]{2012AJ....143..139E} and an aperture correction coefficient $F_\mathrm{corr}=0.91$.\footnote{This so-termed ``maximum scaling factor'' aperture correction coefficient in principle applies to surface brightness measurements of very extended sources. See section~8.2 of \cite{2021..............I}.}

\subsection{\qty{21}{\centi \meter} Line Data} \label{subsec:21cm}

We use the natural-weighted \qty{21}{\centi \meter} line data from \things \cite[]{2008AJ....136.2563W} to trace the atomic hydrogen, \ion{H}{1}. For our sample galaxies, the \qty{21}{\centi \meter} line data typically feature a synthetic beam FWHM of $b_\mathrm{FWHM}\approx \qty{10}{\arcsec}$ or $\sim \qty{400}{\parsec}$. The typical single-channel noise is $\sim \qty{0.4}{\milli \jansky \per \beam}$ or roughly $\sim \qty{0.4}{\solarmass \per \square \parsec}$, assuming a $\sim \qty{20}{\kilo \meter \per \second}$ signal width.

The analyses in the work implicitly assume that the galaxy disks are flat, which is not necessarily the case especially for \ion{H}{1} disks which might warp \cite[]{2007A&A...466..883V}. Indeed, at least three galaxies (\ngc{4736}, \ngc{5055} and \ngc{7793}) are identified to show significant \ion{H}{1} warps within their disk radius $R_{25.5}$, i.e., the radius where the \qty{3.6}{\micro \meter} surface brightness reaches \qty{25.5}{\magnitude \per \square \arsc}, following the method outlined in \cite{2007A&A...468..903J} and \cite{2017MNRAS.472.3029W} using the kinematic information from \cite{2008AJ....136.2648D}. We test and find that removing these three galaxies does not qualitatively alter our major results in Section~\ref{sec:result}.

\subsection{Molecular Gas Line Data and Neutral Gas Distribution} \label{subsec:molecular}

We mainly use the CO $J=2\to 1$ $\nu =\qty{230}{\giga \hertz}$ emission line data from the Heterodyne Receiver Array CO Line Extragalactic Survey \cite[\heracles;][]{2009AJ....137.4670L} to trace the molecular gas. For \ngc{3351}, we use the CO $J=1\to 0$ $\nu =\qty{115}{\giga \hertz}$ emission line data from the \textsc{bima} Survey of Nearby Galaxies \cite[\bimasong;][]{2003ApJS..145..259H}. These molecular line data typically feature a beam FWHM of $b_\mathrm{FWHM}\approx \qty{13}{\arcsec}$ or $\sim \qty{560}{\parsec}$, and reach a depth of $\sim \qty{0.10}{\solarmass \per \square \parsec}$ assuming a $\sim \qty{20}{\kilo \meter \per \second}$ signal width. Neither data are availabe for \ngc{3621}, \ngc{5236} and \ngc{7793}, and we thus estimate the molecular gas distribution by inverting the sub-\qty{}{\kilo \parsec} Schmidt law \cite[]{2008AJ....136.2846B}, with the the star formation rate distribution evaluated using the \qty{15}{\arcsec} resolution [for these three galaxies, $\sim \qty{360}{\parsec}$] $WISE4$ and $FUV$ images retrieved from the $z=0$ Multi-wavelength Galaxy Synthesis database \cite[$z$0\acro{mgs};][]{2019ApJS..244...24L, https://doi.org/10.26131/irsa6}\footnote{\url{https://doi.org/10.26131/IRSA6}}.

We match the beam size and combine the \ion{H}{1} and molecular gas distribution for each galaxies, which are then converted into mass surface density unit following the appendix of the \cite{2009AJ....137.4670L}. Following \cite{2008AJ....136.2782L}, we adopt a uniform working threshold of $\Sigma _{\mathrm{neutral},\, \mathrm{thresh}}=\qty{1}{\solarmass \per \square \parsec}$ in the present work.

\section{Method} \label{sec:method}

\begin{figure*}
    \includegraphics[width=\linewidth]{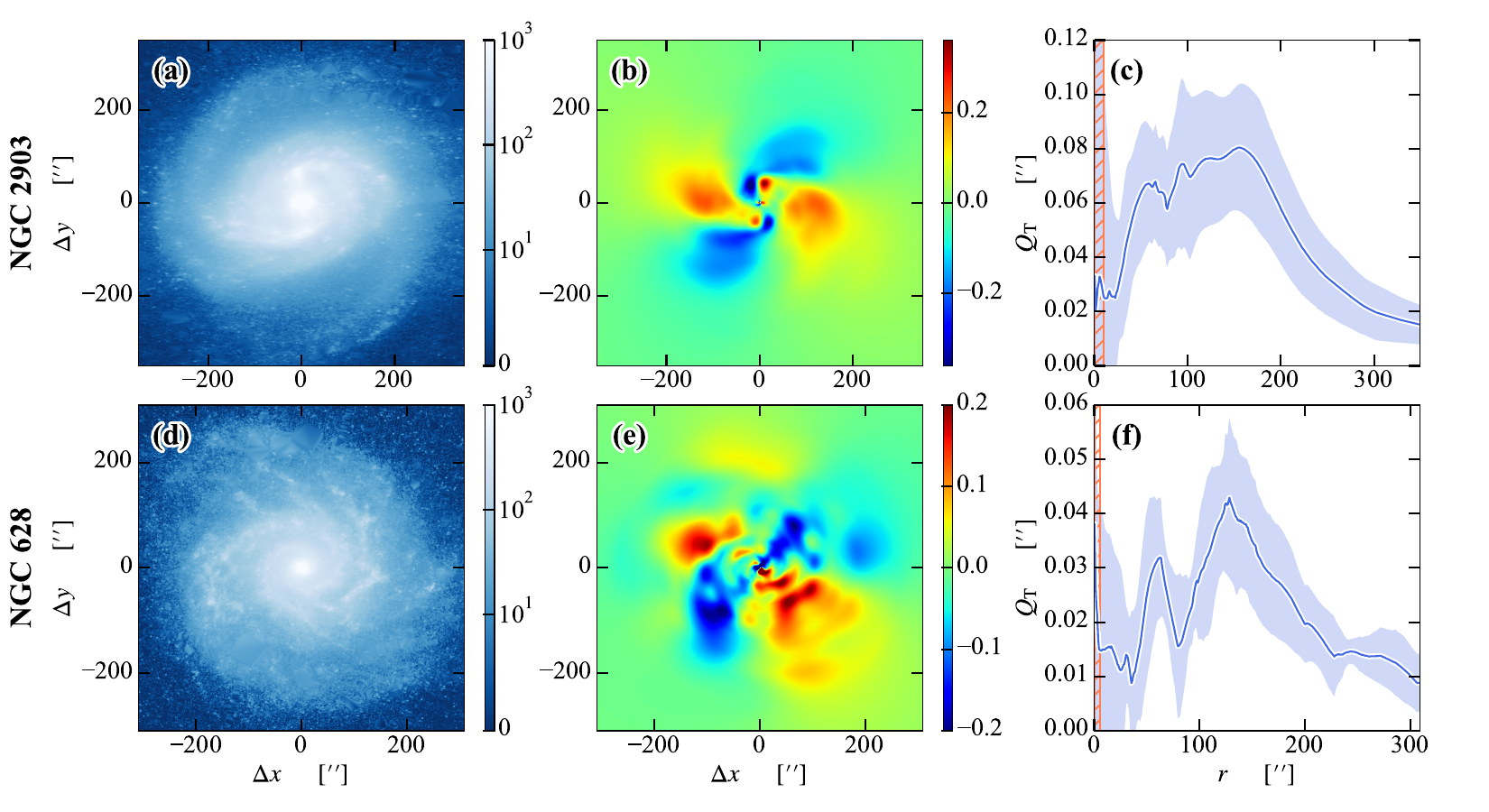}
    \caption{Showcasing torque force field evaluation for barred spiral galaxy \ngc{2903} (top row) and non-barred spiral galaxy \ngc{628} (bottom row). (a, d) De-projected stellar mass surface density in \qty{}{\solarmass \per \square \parsec}. (b, e) Evaluated torque force field $Q_{\mathrm{T}}$, sign being arbitrary. (c, f) Radial profile of $Q_{\mathrm{T}}$. The hatched region show 3 times PSF FWHM.}
    \label{fig:example}
\end{figure*}

We now describe how we characterize the non-axisymmetric structures as well as the neutral gas distribution within the galactic disk. In the present work specifically, we employ optimal bulge stretching correction to improve the accuracy of the torque force field evaluation, and also consider the stabilizing effects of the dark matter halo in addition to the bulge and the symmetric part of the disk (detailed in Section~\ref{subsec:bulge} and Section~\ref{subsec:rc}, respectively).

\subsection{Estimating the Torque Forces} \label{subsec:force}

To quantify the strength of non-axisymmetric structures in terms of perturbing the motion of disk material, we evaluate the tangential force field specifically due to the non-axisymmetric structures \cite[]{1981A&A....96..164C, 1994ApJ...437..162Q}, $f_{\mathrm{t}}$, and the radial force which counters its action, $f_r$. The ratio
\begin{equation} \label{eq:q_st_prof}
    Q_{\mathrm{T}} =\left| f_\mathrm{t}\right| / \left| f_r\right|,
\end{equation}
thus quantifies the impact of the non-axisymmetric structures at each point within the galactic disk \cite[e.g.,][]{1999AJ....117..792S, 2002MNRAS.337.1118L, 2016AAA...587A.160D}.

Typical distributions of the torque force amplitude $Q_{\mathrm{T}}$ are showcased in Figure~\ref{fig:example} for barred galaxy \ngc{2903} and non-barred galaxy \ngc{628}, respectively. From Figure~\ref{fig:example}, one identifies the symbolic four-quadrant, or ``butterfly'' bar perturbation pattern \cite[e.g., inside the bar-dominated region of \ngc{2903},][]{2001ApJ...550..243B, 2002MNRAS.337.1118L}. One also sees that spiral-arms might also induce significant perturbation beyond the bar-dominated region or in non-barred galaxies, which appears more twisted and less regular. Figure~\ref{fig:example} also shows the radial profiles of $Q_{\mathrm{T}}$ by taking azimuthal median at each radius. The multiple peaks in \ngc{2903} correspond to the bar and the spiral arms, respectively, while those in \ngc{628} are exclusively due to spiral arms. In these nearby galaxies, the torque forces are mainly due to the stellar non-axisymmetric structures [blue curves in Figure \ref{fig:example}(c, f)], which is quantified by $Q_{\mathrm{T}}$ in the present work. We also calculate the torque force field with the gas self-gravity combined along with the stellar forces, which is denoted as $Q_{\mathrm{T,\, +gas}}$. Comparison suggests that perturbation due to gaseous features \cite[]{2021ApJ...913..113M, 2023A&A...676A.113S} seems to contribute in a sub-dominant manner (see Section~\ref{subsec:ng} in the Appendix for further discussion), and seems to be less closely related to the neutral gas concentration (Section~\ref{subsec:average}).

In the present work, torque force fields like those shown in Figure \ref{fig:example} are calculated largely following the ``polar'' method \cite[e.g.,][]{1999AJ....117..792S, 2002MNRAS.337.1118L, 2002MNRAS.330...35A, 2003AJ....126.1148B, 2016AAA...587A.160D}. This method integrate the stellar and gas mass surface density field reconstructed from Fourier decomposition
\begin{equation} \label{eq:azimuthal}
    \Sigma _\star \left( r,\,\phi \right) =\Sigma _{\star,\, 0}\left( r \right) \left\{ 1+\sum_{m\in M}{A_m\left( r \right) \cos \left[ m\phi -\phi _m\left( r \right) \right]} \right\} .
\end{equation}
Here, $A_m\left( r \right)$ is the normalized $m$-th order Fourier amplitude and $\phi _m\left( r \right)$ is the corresponding phase angle. Specifically, we choose to consider in Equation~\eqref{eq:azimuthal} $m\in M=\left\{ 0,\, 1,\, 2,\, 3,\, 4,\, 6,\, 8\right\}$ to account for perturbation due to bars \cite[]{1990ApJ...357...71O, 2002MNRAS.337.1118L, 2002MNRAS.330...35A}, various types of spiral arms \cite[]{1988A&AS...76..365C, 2009MNRAS.397.1756D, 2018ApJ...862...13Y}, and disk lopsidedness \cite[][]{2005A&A...438..507B, 2011A&A...530A..30V}. To ensure robust evaluation, we choose not to incorporate higher-order Fourier modes, which are in most cases sub-dominant but might be locally biased by, e.g., low signal-to-noise ratio or dust emission \cite[see Section~\ref{sec:uncertainties} of the Appendix;][]{2002MNRAS.337.1118L, 2016A&A...596A..84D}. We assume a scale height as a function of morphological type $T$ from \cite{1998MNRAS.299..595D} in our evaluation.

The uncertainty of $Q_{\mathrm{T}}$ is mostly propagated from the central position, the disk inclination and position angle, and the uncertain disk scale height. The interested reader is kindly referred to Section~\ref{sec:uncertainties} of the Appendix for a detailed discussion of uncertainty analyses.

We summarize our evaluated torque force field in Section~\ref{sec:individual} of the Appendix for each of the individual galaxies.

\subsection{Improved Bulge Stretching Correction with Optimal Multi-Component Decomposition} \label{subsec:bulge}

Torque force evaluation frequently relies on multi-component decomposition to properly consider the stabilizing effects of the bulge  \cite[]{2004MNRAS.355.1251L, 2016AAA...587A.160D} and avoid artificially ``stretched'' in decomposition. Yet, over-simplified decomposition strategy might lead to inaccurate bulge measurement \cite[]{2009MNRAS.393.1531G, 2017ApJ...845..114G}, thus inaccurate bulge stretching correction to be applied. Optimal bulge decomposition strategy is therefore crucial in yielding robust torque force evaluation, especially in the inner part of the disk.

\begin{figure*}
    \centering
    \includegraphics[width=0.75\linewidth]{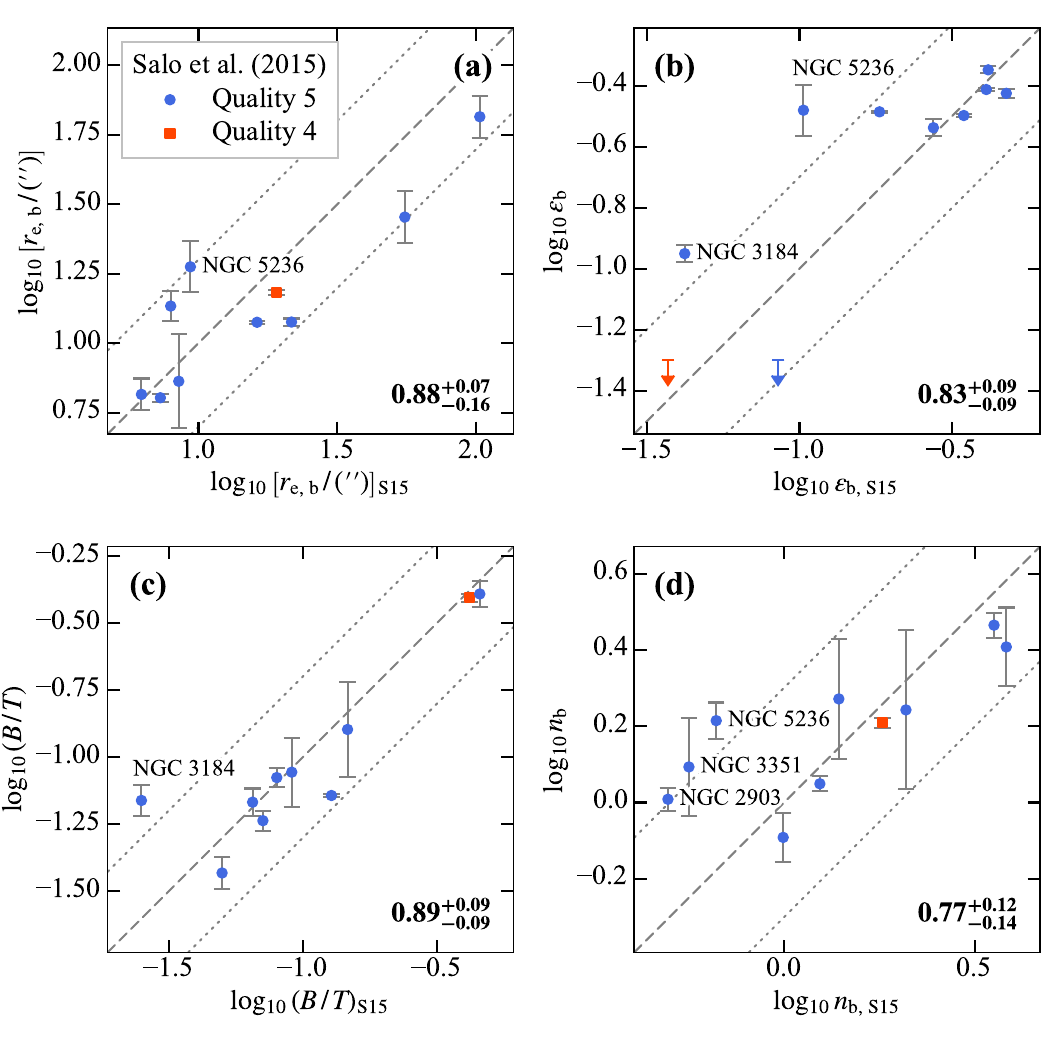}
    \caption{Comparison between the best-fit (a) effective radii $r_{\mathrm{e,\, b}}$, (b) ellipticities $\epsilon _{\mathrm{b}}$, (c) bulge-to-total ratios $B/T$, and (d) Sérsic indices $n_{\mathrm{b}}$, of the bulge components, evaluated from the multi-component decomposition performed in the present work and that presented by \cite{2015APJS..219....4S}. Symbols are differentiated by the quality flags assessed by \cite{2015APJS..219....4S}, where \qty{5}{} (blue circle) represents the most reliable cases and \qty{4}{} (red square) for the less reliable cases. The dashed line and the dotted lines indicate $y=x$ and offsets of $\qty{\pm 0.3}{\dex}$, respectively. Pearson's correlation coefficients are also quoted at the lower right-hand corner in each panel.}
    \label{fig:compare_bulge}
\end{figure*}

We therefore follow the strategy of \cite{2017ApJ...845..114G} to measure bulges in the sample galaxies. \cite{2017ApJ...845..114G} have demonstrated that properly modelling secondary morphological features (such as bars, nuclear bars, nuclear/inner rings, nuclear/inner lenses and disk breaks) is crucial for accurate bulge measurement. We thus choose to construct models including bars, nuclear features, various types of rings and disk breaks, with initial guesses consulting literature dedicated to morphology \cite[]{2015A&A...582A..86H} or from visual inspection.

Figure~\ref{fig:compare_bulge} compares our results with \cite{2015APJS..219....4S}, who have performed multi-component decomposition, yet with a pre-selected, limited number (no more than \qty{4}{}) of components. In comparison, our decomposition does not set a constraint for the number of components, and considers up to \qty{7}{} components for cases with complex morphology. Indeed, one sees cases with significant different effective radii $r_{\mathrm{e,\, b}}$, ellipticities $\epsilon _\mathrm{b}$, bulge-to-total ratios $B/T$, and Sérsic indices $n _{\mathrm{b}}$, in Figure~\ref{fig:compare_bulge}. A further detailed inspection suggests that the differences for the labelled extreme cases could largely be attributed to the difference made by our more dedicated models. We thus consider elaborate multi-component decomposition to be necessary in robust and accurate evaluation of the torque field. The interested reader is referred to Section~\ref{subsec:comparison_decompose} of the Appendix for this closer inspection.

We follow the formulae in \cite{2020A&A...635A..20V} to estimate the radial force from the bulges measurements returned by our optimal multi-component decomposition. Note that \cite{2020A&A...635A..20V} assumes spherical geometry, which is presumably not a good approximation for disk-like pseudo-bulges \cite[]{2004ARA&A..42..603K, 2005MNRAS.358.1477A, 2020A&A...643A..14G}. We test and find that the uncertainty introduced by the assumed geometry is sub-dominant compared to what would have been introduced by the bulge stretching effects, mitigating the concern to some extent (Section~\ref{sec:uncertainties} of the Appendix).

\subsection{Improved Estimation for Total Radial Force} \label{subsec:rc}

Alongside the bulge and the symmetric part of disk, dark matter halo arguably also stabilizes the disk and counters the action of the non-axisymmetric structures, especially at large radii \cite[]{2008AJ....136.2648D}. It is possible to globally account for this effects, e.g., using \ion{H}{1} line width \cite[]{2016AAA...587A.160D}. One might also wish to use the \ion{H}{1} rotation curve to construct a more detailed estimation for the radial force at each radius, yet accurately measuring the rotation curve is itself challenging if there are strong non-axisymmetric structures and thus strong streaming motion \cite[]{2008MNRAS.385..553D, 2015MNRAS.454.3743R, 2018A&A...618A.106R}. Most importantly, both methods are impractical for galaxy with low inclination, which are the most ideal targets for characterizing the non-axisymmetric structures yet the difficult cases to extract reliable kinematic information.

In the present work, alternatively, we make use of the gravitational acceleration relation to account for the stabilizing effects of the dark matter halo. The gravitational acceleration relation \cite[GAR;][]{2017MNRAS.468L..68L, 2017ApJ...836..152L} is an observational empirical relation between the radial force solely due to baryonic matter (i.e., stars and gas), and that due to dark matter. GAR is remarkably tight with scatter of a mere \qtyrange[range-phrase=--, range-units=bracket]{0.13}{0.15}{dex} \cite[]{2017ApJ...836..152L}, making it a powerful tool to estimate the total radial force at each location within the disk. The GAR parameters from \cite{2017ApJ...836..152L} are adopted. When the \ion{H}{1} rotation curve is available from \cite{2008AJ....136.2648D}, we scales the estimated total radial force to match the measured \ion{H}{1} rotation curve \cite[techically, by fitting for a free $\hat y$ in equation~9 of][]{2017ApJ...836..152L} to further improve the accuracy slightly.

\section{Results} \label{sec:result}


\begin{deluxetable*}{cccccc}
    \tabletypesize{\footnotesize}
    \tablewidth{0pt}
    \tablecaption{Summary of Correlation Coefficients}
    \label{table:corr}
    \tablehead{
        \colhead{Index} &
        \colhead{$x$} &
        \colhead{Spearman's Correlation Coefficient} &
        \multicolumn{2}{c}{Partial Spearman's Correlation Coefficient} &
        \colhead{Note} \\
        \cline{4-5}
        \colhead{} &
        \colhead{} &
        \colhead{} &
        \colhead{$\log _{10}\left( \bar \Sigma _{\star}/\qty{}{\solarmass \per \square \parsec}\right)$} &
        \colhead{Neutral Gas Fraction} &
        \colhead{}
    }
    \colnumbers
    \startdata
    \multicolumn{6}{c}{Correlations with $C_{\mathrm{neutral}}$} \\
    \hline
    1 & $\log _{10}\bar Q_{\mathrm{T}}$ & $\mathbf{\qtyerrs{0.56}{0.16}{0.21}}$ ($\mathbf{\qtyerrs{0.02}{0.15}{0.02}}$) & $\mathbf{\qtyerrs{0.61}{0.14}{0.28}}$ ($\mathbf{\qtyerrs{0.01}{0.18}{0.01}}$) & $\mathbf{\qtyerrs{0.54}{0.17}{0.41}}$ ($\mathbf{\qtyerrs{0.03}{0.48}{0.03}}$) &  \\
    2 & $\log _{10}\bar Q_{\mathrm{T,\, +gas}}$ & $\qtyerrs{0.39}{0.23}{0.25}$ ($\qtyerrs{0.12}{0.41}{0.11}$) & $\mathbf{\qtyerrs{0.51}{0.18}{0.29}}$ ($\mathbf{\qtyerrs{0.04}{0.27}{0.04}}$) & $\qtyerrs{0.42}{0.21}{0.40}$ ($\qtyerrs{0.11}{0.49}{0.10}$) & * \\
    3 & $2+\log _{10}s_{\mathrm{disk}}$ & $\qtyerrs{0.45}{0.20}{0.22}$ ($\qtyerrs{0.07}{0.31}{0.07}$) & $\qtyerrs{0.47}{0.17}{0.24}$ ($\qtyerrs{0.07}{0.31}{0.06}$) & $\qtyerrs{0.43}{0.18}{0.29}$ ($\qtyerrs{0.09}{0.39}{0.08}$) &  \\
    4 & $\log _{10}Q_{\mathrm{g}}$ & $\mathbf{\qtyerrs{0.76}{0.11}{0.18}}$ ($\mathbf{\qtyerrs{0.00}{0.02}{0.00}}$) & $\mathbf{\qtyerrs{0.72}{0.13}{0.38}}$ ($\mathbf{\qtyerrs{0.00}{0.22}{0.00}}$) & $\mathbf{\qtyerrs{0.68}{0.16}{0.63}}$ ($\mathbf{\qtyerrs{0.01}{0.48}{0.01}}$) & $\dagger$ \\
    5 & $\log _{10}Q_{\mathrm{g,\, +gas}}$ & $\mathbf{\qtyerrs{0.61}{0.16}{0.24}}$ ($\mathbf{\qtyerrs{0.01}{0.16}{0.01}}$) & $\mathbf{\qtyerrs{0.64}{0.15}{0.35}}$ ($\mathbf{\qtyerrs{0.01}{0.28}{0.01}}$) & $\mathbf{\qtyerrs{0.54}{0.21}{0.57}}$ ($\mathbf{\qtyerrs{0.05}{0.56}{0.04}}$) & *, $\dagger$ \\
    6 & $\log _{10}\bar \Gamma$ & $\qtyerrs{0.33}{0.23}{0.24}$ ($\qtyerrs{0.20}{0.43}{0.18}$) & $\mathbf{\qtyerrs{0.50}{0.16}{0.25}}$ ($\mathbf{\qtyerrs{0.05}{0.28}{0.04}}$) & $\qtyerrs{0.40}{0.19}{0.40}$ ($\qtyerrs{0.13}{0.49}{0.11}$) &  \\
    7 & $\log _{10}\bar \Gamma _{\mathrm{+gas}}$ & $\qtyerrs{0.15}{0.27}{0.27}$ ($\qtyerrs{0.56}{0.24}{0.48}$) & $\qtyerrs{0.41}{0.16}{0.25}$ ($\qtyerrs{0.12}{0.35}{0.10}$) & $\qtyerrs{0.33}{0.18}{0.33}$ ($\qtyerrs{0.21}{0.47}{0.17}$) & * \\
    \hline
    \multicolumn{6}{c}{Correlations with $C_{\mathrm{CO}}$} \\
    \hline
    8 & $\log _{10}\bar Q_{\mathrm{T}}$ & $\mathbf{\qtyerrs{0.51}{0.26}{0.28}}$ ($\mathbf{\qtyerrs{0.04}{0.32}{0.04}}$) & $\mathbf{\qtyerrs{0.50}{0.26}{0.33}}$ ($\mathbf{\qtyerrs{0.05}{0.40}{0.05}}$) & $\mathbf{\qtyerrs{0.51}{0.26}{0.30}}$ ($\mathbf{\qtyerrs{0.04}{0.36}{0.04}}$) &  \\
    9 & $\log _{10}\bar Q_{\mathrm{T,\, +gas}}$ & $\qtyerrs{0.42}{0.25}{0.28}$ ($\qtyerrs{0.09}{0.43}{0.09}$) & $\qtyerrs{0.45}{0.23}{0.33}$ ($\qtyerrs{0.08}{0.45}{0.08}$) & $\qtyerrs{0.43}{0.25}{0.31}$ ($\qtyerrs{0.10}{0.45}{0.10}$) & * \\
    10 & $2+\log _{10}s_{\mathrm{disk}}$ & $\qtyerrs{0.42}{0.24}{0.26}$ ($\qtyerrs{0.09}{0.39}{0.09}$) & $\qtyerrs{0.41}{0.24}{0.31}$ ($\qtyerrs{0.11}{0.46}{0.11}$) & $\qtyerrs{0.43}{0.23}{0.29}$ ($\qtyerrs{0.10}{0.43}{0.09}$) &  \\
    \enddata
    \tablecomments{
        Summary of correlation coefficients. Brackted values are the corresponding $p$-values. Uncertainties are quoted as \qty{16}{\percent} and \qty{84}{\percent} percentiles evaluated from the bootstrap sampling method. Statistically signifiant correlations (with $p<\qty{0.05}{}$ at face values) are highlighted in bold typeface. (1) Index. (2) Abscissa $x$. (3) Spearman's correlation coefficient. (4) Partial Spearman's correlation coefficient, controlling for the average stellar mass surface density $\log _{10}\left( \bar \Sigma _{\star}/\qty{}{\solarmass \per \square \parsec}\right)$. (5) Partial Spearman's correlation coefficient, controlling for the neutral gas fraction within the corresponding radial range. For correlations with $C_{\mathrm{neutral}}$ this is $f_{\mathrm{neutral}}\left( R_{25.5}\right)$, for correlations with $C_{\mathrm{CO}}$ this is $f_{\mathrm{neutral}}\left( r_{50}\right)$. (6) Note. \\
        * Gas self-gravity is considered in evaluating the torque field, and the relevant parameters. \\
        $\dagger$ Only galaxies with $Q_{\mathrm{g}}$ and $Q_{\mathrm{g,\, +gas}}$ reliably determined from the torque forces radial profiles are considered (Section \ref{subsec:maximal}).
    }
\end{deluxetable*}


We are now in a position to investigate how the neutral gas distribution is linked to the \textit{strength} of non-axisymmetric structures, quantitatively measured as the torque forces they exert.

\subsection{Neutral Gas Concentration and Average Torque Forces} \label{subsec:average}

\begin{figure}
    \centering
    \includegraphics[width=\columnwidth]{"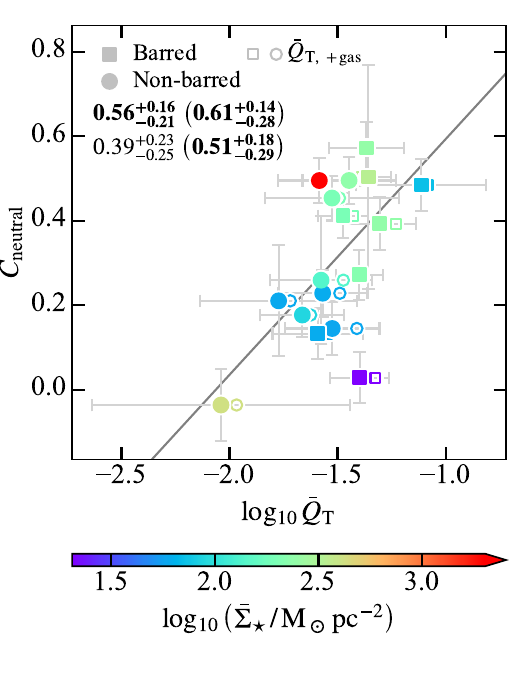"}
    \caption{Correlations between the average torque forces 
    $\bar Q_{\mathrm{T}}$ and the inner neutral gas concentration parameter $C_{\mathrm{neutral}}$. Symbols are color-coded by the average stellar surface density, $\bar \Sigma _{\star}$, and are distinguished for barred (square) and non-barred (circle) galaxies. Open symbols represent results for $\bar Q_{\mathrm{T,\, +gas}}$, i.e., when gas self-gravity is combined in evaluating the torque forces.  Spearman's and partial Spearman's correlation coefficients (controlling for $\bar \Sigma _{\star}$, bracketed) are quoted at the upper left-hand corner of the figure for $\bar Q_{\mathrm{T}}$ (first row) and $\bar Q_{\mathrm{T,\, +gas}}$ (second row), respectively. Statistically significant ($p<\qty{0.05}{}$) correlations are highlighted in bold typeface.}
    \label{fig:result:im_c05}
\end{figure}

We begin by investigating how the neutral gas radial distribution might be related to the torque field exerted by the non-axisymmetric structures.

Figure~\ref{fig:neutral_gas_profile} serves as a quick-look table for neutral gas radial profiles of all galaxies sorted by the \textit{radially average torque forces within the disk} exerted by the stellar non-axisymmetric structures,
\begin{equation} \label{eq:average_torque}
    \bar{Q}_{\mathrm{T}}=\mathrm{average}\left\{ Q_{\mathrm{T}}\left( r \right) \right\} ,\qquad 3\gamma ^\prime_{\mathrm{PSF}}<r<R_{25.5}.
\end{equation}
Here, the $\gamma ^\prime_{\mathrm{PSF}}$ refers to the PSF FWHM projected onto the galaxy disk plane. We define the \textit{inner neutral gas concentration parameter}
\begin{equation}
    C_{\mathrm{neutral}}=\log _{10}\left[ \frac{M_{\mathrm{neutral}}\left( 0.5R_{25.5} \right)}{M_{\mathrm{neutral}}\left( R_{25.5} \right)} \right] +\mathrm{constant},
\end{equation}
to capture this disk-scale concentration of neutral gas. Though this definition of $C_{\mathrm{neutral}}$ appears similar to those proposed by \cite{2007PASJ...59..117K} and \cite{2022AaA...666A.175Y}, we emphasize that the concentration parameters defined by these authors, comparatively, highlights the central \qty{1}{\kilo \parsec} region instead of the disk-scale.

Figure~\ref{fig:result:im_c05} shows that there is a positive correlation between $\bar{Q}_{\mathrm{T}}$ and $C_{\mathrm{neutral}}$, despite the small size of the sample. Furthermore, one sees that barred galaxies (square) and non-barred galaxies (circle) form a continuous sequence together in this $\bar Q_{\mathrm{T}}$--$C_{\mathrm{neutral}}$ parameter space. To our knowledge, this continuous sequence in $\bar{Q}_{\mathrm{T}}$ and $C_{\mathrm{neutral}}$ have remained elusive in previous works, both in the sense that the torque forces $\bar{Q}_{\mathrm{T}}$ directly quantify the physical effects of non-axisymmetric structures, and that $C_{\mathrm{neutral}}$ refers to neutral gas distribution in a relatively wide disk radial range.

All sample galaxies taken together, we find a Spearman's correlation coefficient of $\rho=\qtyerrs{0.56}{0.16}{0.21}$, with a corresponding $p$-value $p=\qtyerrs{0.02}{0.15}{0.02}$ between $\bar Q_{\mathrm{T}}$ and $C_{\mathrm{neutral}}$, indicating relatively strong, statistically significant correlation, though a rigorous conclusion is precluded due to the small sample size. Interestingly, when gas self-gravity is considered in evaluating the torque forces, the correlation between $\bar Q_{\mathrm{T,\, +gas}}$ and $C_{\mathrm{neutral}}$ is seen to be weaker and statistically insignificant (See Table~\ref{table:corr} for a summary of statistics). This seems to suggest that the torque forces exerted by stars are more closely related to the neutral gas distribution within the galactic disk, while gas self-gravity does not contribute significantly \cite[]{2022MNRAS.512.1522D}. Another possibility is that the non-axisymmetric part of the self-gravity forces might be significantly contributed by shock fronts, shells, bubbles and holes of sizes up to $\sim \qty{1}{\kilo \parsec}$, driven by spiral arm dynamics and stellar feedback, and further stretched and elongated by the shearing motion \cite[e.g.,][]{1990A&A...227..175P, 2009ApJ...704.1538W, 2019ApJ...872....5S, 2023ApJ...944L..24W, 2023ApJ...944L..22B}. While playing an important role in shaping the overall disk cold gas structures, some of these features could be transient \cite[e.g., typical lifetimes of shell-like features $<\qty{10}{\mega \yr}$,][]{2023ApJ...944L..24W}, and might thus be difficult to have a coherent impact on the radial distribution of disk neutral gas, which is expected on timescales of the disk dynamical timescale ($\sim \qty{100}{\mega \yr}$). Here and in what follows, the uncertainties of statistics are evaluated by bootstrapping the sample to account for the dominant impact of the small sample size, and are quoted as the \qty{16}{\percent} and the \qty{84}{\percent} percentiles, respectively.

\begin{figure*}
    \centering
    \includegraphics[width=0.75\linewidth]{"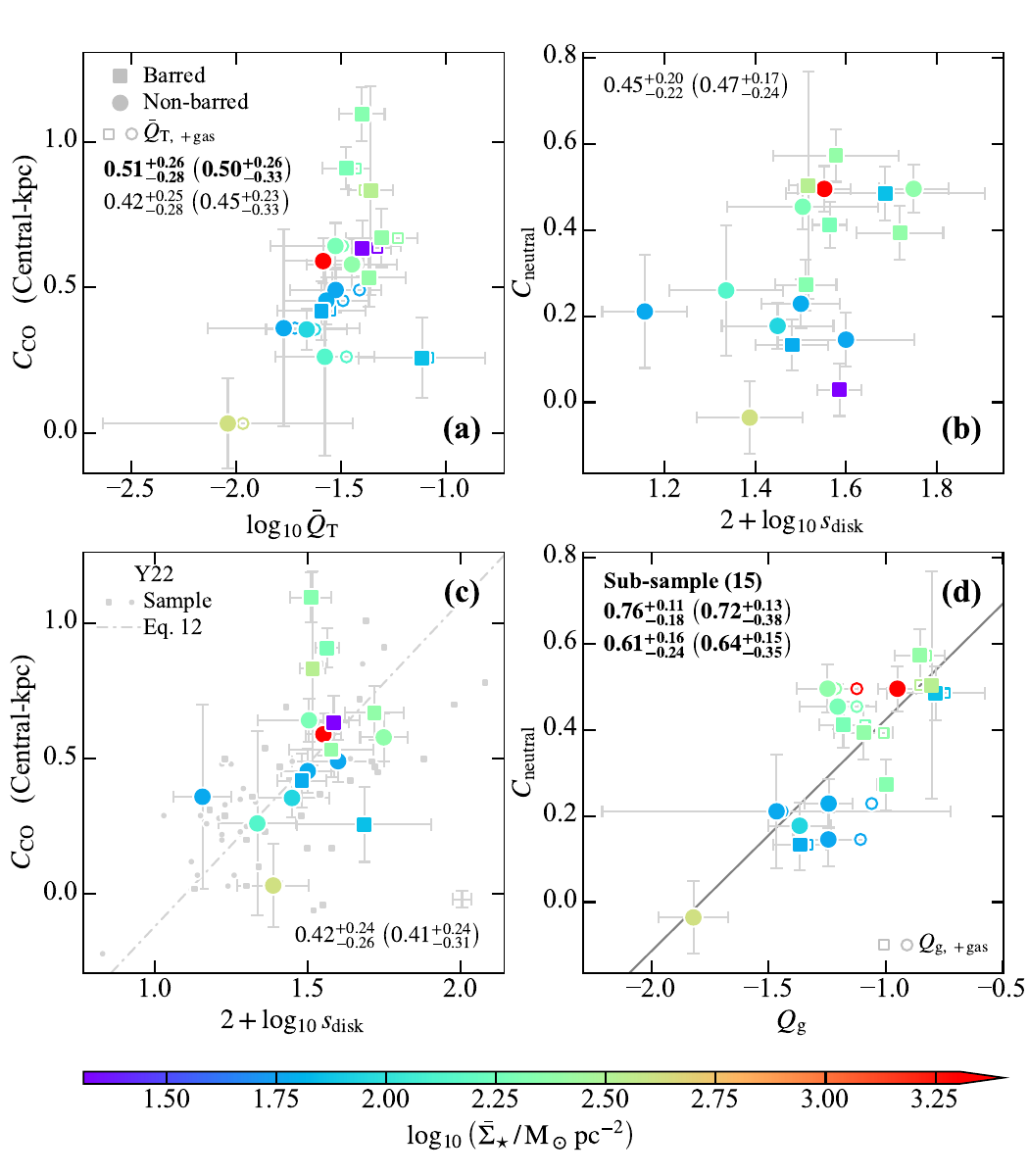"}
    \caption{The neutral and molecular gas concentration and non-axisymmetric structures strength, quantified by different measures. Symbols style and color are the same to Figure~\ref{fig:result:im_c05}. Spearman's and partial correlation coefficients (controlling for $\bar \Sigma _{\star}$, bracketed) are quoted at the corner of each panel. In panels (a, d), correlation coefficients for torque forces evaluated without (first row) and with gas self-gravity considered (second row) are quoted, respectively. Statistically significant ($p<\qty{0.05}{}$) correlations are highlighted in bold typeface. (a) The radially average torque forces $\bar Q_{\mathrm{T}}$ and the central \qty{1}{\kilo \parsec} molecular gas concentration index $C_{\mathrm{CO}}$. (b) The average non-axisymmetric surface density Fourier amplitudes $2+\log _{10}s_{\mathrm{disk}}$ and the inner neutral gas concentration parameter $C_{\mathrm{neutral}}$. (c) $2+\log _{10}s_{\mathrm{disk}}$ and $C_{\mathrm{CO}}$. Measurements (smaller symbols) and linear regression (grey line) from \cite{2022AaA...666A.175Y} are also shown. (d) The maximal torque forces $Q_{\mathrm{g}}$ and $C_{\mathrm{neutral}}$. Marginally strong and significant ($p\lesssim \qty{0.05}{}$) correlations could be seen in (a) for stellar $\bar Q_{\mathrm{T}}$, no strong and statistically significant correlation is seen in (b, c) for our sample, while strong and statistically significant correlations are seen in (d).}
    \label{fig:result:corrs}
\end{figure*}

It seems most likely that the correlation seen in Figure~\ref{fig:result:im_c05} is reflecting gas concentration \textit{on the disk-scale}, instead of simply because the gas in the central \qty{1}{\kilo \parsec} is enhanced \cite[e.g.,][]{1999ApJ...525..691S, 2005ApJ...632..217S, 2007PASJ...59..117K, 2012MNRAS.423.3486W, 2019MNRAS.484.5192C}. For a comparison, we present the average torque forces $\bar Q_{\mathrm{T}}$ and the central-\qty{}{\kilo \parsec} molecular gas concentration index $C_{\mathrm{CO}}$ \cite[defined in][]{2022AaA...666A.175Y} in Figure~\ref{fig:result:corrs}(a). The correlation between $\bar Q_{\mathrm{T}}$ and $C_{\mathrm{CO}}$ is almost as strong as that between $\bar Q_{\mathrm{T}}$ and $C_{\mathrm{neutral}}$, but becomes weaker than the latter when we consider partial correlations in the next sub-section. Furthermore, although we do not show explicitly, if we mimic the definition of $C_{\mathrm{CO}}$ \cite[]{2022AaA...666A.175Y} to define a ``central-\qty{}{kpc} neutral gas concentration index'' considering both molecular and atomic gas in the central-\qty{}{\kilo \parsec} regions, then the correlation becomes even weaker and less statistically significant. It implies that processes other than neutral gas radial transportation might also be at play, like the conversion of \ion{H}{1} to the molecular gas, to display the strong $\bar Q_{\mathrm{T}}$-$C_{\mathrm{CO}}$ relation. The relatively safe conclusion here is that, the disk-scale $\bar Q_{\mathrm{T}}$-$C_{\mathrm{neutral}}$ relation is unlikely caused by the \qty{1}{\kilo \parsec}-scale $\bar Q_{\mathrm{T}}$-$C_{\mathrm{CO}}$ one, as otherwise the former should be weaker than the latter. Instead, the $\bar Q_{\mathrm{T}}$-$C_{\mathrm{neutral}}$ relation is marginally stronger (with $\sim \qty{0.2}{}\sigma$ significance in difference of Spearman's $\rho$), which is physically reasonable as non-axisymmetric structures spread disk-wide and $\bar Q_{\mathrm{T}}$ is a disk-scale measure.

\subsection{Partial Correlations} \label{subsec:partial}

We investigate whether the correlation discussed above holds if we take covariant into consideration (i.e., control for mutually dependent parameter), most relevantly the neutral gas fraction within the corresponding radial range and the average stellar mass surface density. This is important as part of our conclusions, since it has been known that there is a complex interplay between the gas concentration, the non-axisymmetric structures, and the aforementioned factors. The gas content affects the gas concentration \cite[e.g., self-regulation of \ion{H}{1} as is revealed by the size-mass relation,][]{2016MNRAS.460.2143W, 2020ApJ...890...63W} and the strengths of the non-axisymmetric structures \cite[]{2012MNRAS.424.2180M}, simultaneously. Likewise, the stellar mass surface density might also affect the strengths of the non-axisymmetric structures \cite[]{2017MNRAS.471.1070B}, and the transition between \ion{H}{1} and molecular gas to affects the self-regulation of gas \cite[]{2009AJ....137.4670L, 2010ApJ...721..975O, 2011MNRAS.418.1649L}. This complex interplay incompletely outlined above entails an inspection of partial correlations before one infers causality.

We find that our correlation between $\bar Q_{\mathrm{T}}$ and $C_{\mathrm{neutral}}$ is unlikely due to the aforementioned mutual dependence on the stellar mass surface density or the neutral gas fraction. Instead, controlling for the mutually dependent parameters might enhance the correlation. In Figure~\ref{fig:result:im_c05}, symbols are color-coded by the average stellar mass surface density
\begin{equation}
    \bar \Sigma _\mathrm{\star} = \frac{0.5M_\star}{\mathrm{\pi} r^2_{50}}.
\end{equation}
Comparison of Spearman's correlation coefficients and the corresponding $p$-values suggests that the correlation between $\bar Q_{\mathrm{T}}$ and $C_{\mathrm{neutral}}$ is unlikely originated from neutral gas distribution and non-axisymmetric structures being simultaneously affected by $\bar \Sigma _\mathrm{\star}$. Instead, it is interesting to see that the extreme $\bar \Sigma _\mathrm{\star}$ seems to explain the two sample galaxies that show the largest offsets from the linear regression presented in Figure~\ref{fig:result:im_c05}, seemingly reflecting the interplay between stellar mass surface density and gas \cite[]{2009AJ....137.4670L, 2010ApJ...721..975O, 2011MNRAS.418.1649L} in these extreme cases. As a result, statistics suggest that the correlation between $\bar Q_{\mathrm{T}}$ and $C_{\mathrm{neutral}}$ seems to become stronger as $\bar \Sigma _\mathrm{\star}$ is controlled for. The partial correlation between $\bar Q_{\mathrm{T,\, +gas}}$ and $C_{\mathrm{neutral}}$ also becomes statistically significant when controlling for $\bar \Sigma _\mathrm{\star}$, yet the correlation is comparatively weaker than the one not including the gas self-gravity, again suggesting the important role of torque forces induced by stars rather than the neutral gas self-gravity. The correlation seems to be less dependent on the gas fraction, here defined as
\begin{equation}
    f_{\mathrm{neutral}}\left( R_{25.5} \right) =\frac{M_{\mathrm{neutral}}\left( R_{25.5} \right)}{M_{\star}\left( R_{25.5} \right)},
\end{equation}
yet the partial correlation controlling for $f_{\mathrm{neutral}}\left( R_{25.5} \right)$ remains statistically significant. See Section~\ref{sec:partial_gas_fraction} in the Appendix for figures color-coded by the gas fraction in the corresponding radial range instead.

In contrast, the partial correlations between $\bar Q_{\mathrm{T}}$ and $C_{\mathrm{CO}}$ are consistently weaker and marginally significant at best. No statistically signifiant correlation between $\bar Q_{\mathrm{T,\, +gas}}$ and $C_{\mathrm{CO}}$ could be seen. These results further strengthen our confidence that the correlation seen in Figure~\ref{fig:result:im_c05} is not likely simply due to the enhanced molecular gas content in the central-\qty{}{\kilo \parsec} regions.

\subsection{Non-Axisymmetric Structures Torque Forces versus Surface Density Fourier Amplitudes} \label{subsec:fourier}

\cite{2022AaA...666A.175Y} have quantified the strength of non-axisymmetric structures by the the surface density Fourier amplitudes $2+\log _{10}s_{\mathrm{disk}}$, where
\begin{equation} \label{eq:yu22}
    s_{\mathrm{disk}}=\mathrm{average}\left\{ \sqrt{A_{2}^{2}\left( r \right) +A_{3}^{2}\left( r \right) +A_{4}^{2}\left( r \right)} \right\} ,
\end{equation}
with $A_m$ defined in Equation~\eqref{eq:azimuthal} evaluated within the disk-dominated regions \cite[for detailed definition, see][]{2018ApJ...862...13Y, 2022AaA...666A.175Y}. These authors have shown that $C_{\mathrm{CO}}$ correlates with $2+\log _{10}s_{\mathrm{disk}}$ for both barred and non-barred galaxies. We make comparison to understand which measure better quantifies the efficiency of non-axisymmetric structures in affecting the neutral gas distribution.

Figure~\ref{fig:result:corrs}(b) suggests that there is no statistically significant correlation between $2+\log _{10}{s_{\mathrm{disk}}}$ and $C_{\mathrm{neutral}}$ for our sample. Although a positive trend could be vaguely seen, Spearman's statistics suggest that our small sample does not support a strong and statistical significant correlation between $s_{\mathrm{disk}}$ and $C_{\mathrm{neutral}}$ (Table~\ref{table:corr}). Controlling for $\bar \Sigma_{\star}$ or $f_{\mathrm{neutral}}\left( R_{25.5}\right)$, the partial correlation strength between $2+\log _{10}{s_{\mathrm{disk}}}$ and $C_{\mathrm{neutral}}$ and the statistical significance remain largely unchanged. It therefore seems that the correlation between $\bar Q_{\mathrm{T}}$ and $C_{\mathrm{neutral}}$ is more intrinsic.

In Figure~\ref{fig:result:corrs}(c), we show $2+\log _{10}s_{\mathrm{disk}}$ and the central-\qty{}{\kilo \parsec} molecular gas concentration index $C_{\mathrm{CO}}$. We reach similar conclusions as in Figure~\ref{fig:result:corrs}(b). It is to be noted that our sample majorly consists of galaxies with intermediate strength of non-axisymmetric structures, in contrast with that of \cite{2022AaA...666A.175Y} [underlain in Figure~\ref{fig:result:corrs}(c)]. Our point is that, it seems that the radially average torque forces parameter $\bar Q_{\mathrm{T}}$ seems to sensitively quantify the effect of neutral gas distribution being regulated by the non-axisymmetric structures, even when the $s_{\mathrm{disk}}$ dynamical range is limited.

\subsection{Neutral Gas Distribution and the Maximal Torque Forces} \label{subsec:maximal}

We further investigate how the neutral gas distribution might be correlated with the \textit{maximal} torque forces \cite[e.g.][]{2002MNRAS.337.1118L, 2004AJ....127..279B, 2010AJ....139.2465G} 
\begin{equation}
    Q_{\mathrm{g}}=\max \left\{ Q_{\mathrm{T}}\left( r \right) \right\} ,\qquad 3\gamma _{\mathrm{PSF}}^{\prime}<r<R_{25.5},
\end{equation}
and similarly also $Q_{\mathrm{g,\, +gas}}$, within the galaxy disk. Compared to the average counterpart, this maximal torque forces arguably highlight more specifically contribution from the strongest component, and measure the utmost extent to which the motion of neutral gas could be affected. In this regard, it would also be interesting to examine whether the maximal torque forces correlate with the neutral gas concentration as well as neutral gas over-densities at smaller scales.

Before we move on to show the results, it is to be noted that $Q_{\mathrm{g}}$ and $Q_{\mathrm{g,\, +gas}}$ could not always be precisely measured, especially in cases of late-type galaxies. In our sample, for \ngc{925} and \ngc{3621}, specifically, we observe torque forces decreasing almost monotonically with radius, thus a maximal torque force would occur at $r\to 0$. This is neither physical nor reliable, since the central pixels are inevitably affected by the (intrinsically asymmetric and artificially elongated in de-projection) shape of the PSF. We therefore refrain from determining the maximal torque forces for these two galaxies, and omit them in the analyses below in this Section. See Section~\ref{sec:individual} in the Appendix for torque forces $Q_{\mathrm{T}}\left( r\right)$ measured for each individual galaxy.

\subsubsection{The Disk-Scale Neutral Gas Concentration} \label{subsec:qg_concentration}

With this caveat in mind, we see in Figure~\ref{fig:result:corrs}(d) that there is also a positive correlation between the maximal torque forces $Q_{\mathrm{g}}$ and the inner neutral gas concentration parameter $C_{\mathrm{neutral}}$. With a reduced sub-sample, it seems that this correlation is strong and statistically significant despite the small sample-size. The partial correlations controlling for $\bar \Sigma _\mathrm{\star}$ or $f_{\mathrm{neutral}}\left( R_{25.5} \right)$ are also seen to be strong and statistically significant. A positive correlation between $Q_{\mathrm{g,\, +gas}}$ and $C_{\mathrm{neutral}}$ could also be seen, though the statistical significance is marginal when controlling for $f_{\mathrm{neutral}}\left( R_{25.5} \right)$.

It would thus be interesting to examine which of the correlations, i.e., whether it is the $\bar Q_{\mathrm{T}}$--$C_{\mathrm{neutral}}$ correlation (Figure~\ref{fig:result:im_c05}) or the $Q_{\mathrm{g}}$--$C_{\mathrm{neutral}}$ correlation [Figure~\ref{fig:result:corrs}(d)], is more intrinsic to provide constraints on the detailed underlying physics. However, although the correlation coefficients and the $p$-values appear to hint towards stronger correlations with the latter, the statistics are derived over the \textit{reduced sub-sample} for the $Q_{\mathrm{g}}$--$C_{\mathrm{neutral}}$ correlation, and when we force the sample to be the same, the Spearman's correlation coefficients and the corresponding $p$-values become quite close. Considering the statistical uncertainties, we therefore refrain from reaching a solid conclusion, and leave this comparison to further investigations where larger samples are accessible.

\subsubsection{Localized Neutral Gas Over-Densities} \label{subsec:qg_peaks}

\begin{figure*}
    \centering
    \includegraphics[width=\linewidth]{"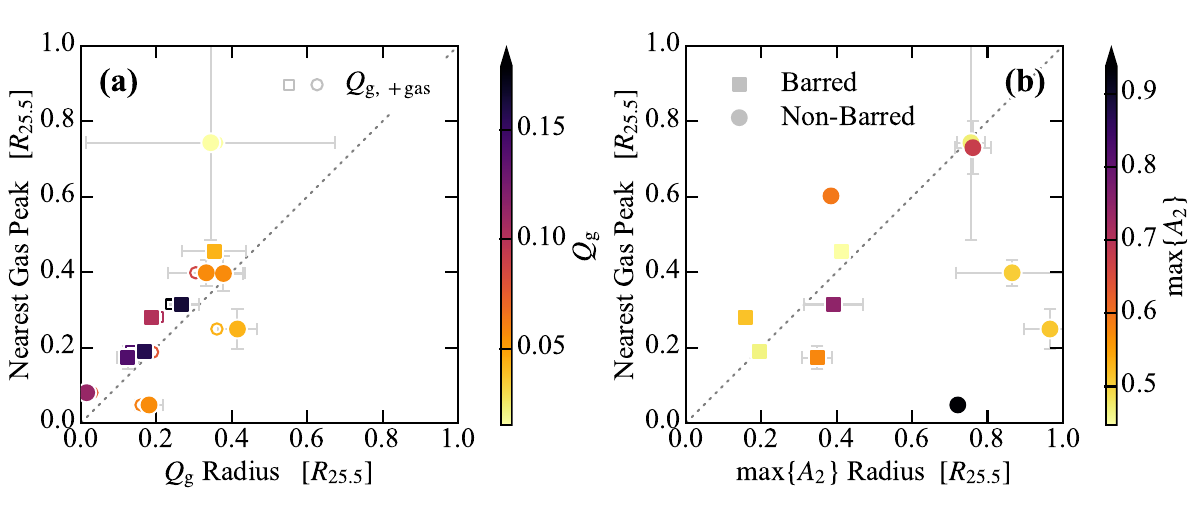"}
    \caption{(a) Radius of maximal torque forces $Q_{\mathrm{g}}$ plotted against the radius of the nearest \qty{}{\kilo \parsec}-scale neutral gas overdensity. Symbols are colored by the corresponding $Q_{\mathrm{g}}$ ($Q_{\mathrm{g,\, +gas}}$), and are distinguished for barred (square) and non-barred (circle) galaxies. Open symbols show the results if we consider the gas self-gravity in evaluating the maximal torque forces $Q_{\mathrm{g,\, +gas}}$. (b) Radius of maximal $m=2$ surface density amplitude $\max \left\{ A_2\right\}$ and radius of the nearest gas overdensity, with symbols colored by the corresponding $\max \left\{ A_2\right\}$. It seems that the \qty{}{\kilo \parsec}-scale gas over-density tends to occur in the vicinity of the maximal torque forces, especially when the torque forces are comparatively strong ($Q_{\mathrm{g}}\gtrsim \qty{0.05}{}$), but less likely for $\max \left\{ A_2\right\}$.}
    \label{fig:result:peaks}
\end{figure*}

Lastly, we divert our attention to the \qty{}{\kilo \parsec}-scale neutral gas over-densities readily seen from the radial profiles. The \texttt{find\_peak} function in \textsc{SciPy.signal} module is used to extract these over-densities as local peaks from the neutral gas radial profiles within $R_{25.5}$. It turns out that the width of these peaks are typically $\sim \qty{1}{\kilo \parsec}$, hence the terminology. \ngc{2903}, \ngc{5055}, \ngc{6946}, and \ngc{7793} feature generally smooth neutral gas radial profiles from which no local peak could be identified, and are thus omitted from the analyses below in this Section. A usual prediction from theories and simulations is that these dissipative processes could be responsible to accumulate the gas at the shock front \cite[e.g.,][]{1992MNRAS.259..345A, 2010ApJ...715L..56S,2014MNRAS.440..208K}. In line with these predictions, since gas motion is perturbed most strongly near the location of maximal torque force, it is natural for us to conjecture that there is a connection between the maximal torque force and the \qty{}{\kilo \parsec}-scale gas over-densities. Below, we look into this conjecture with our measurements.

In Figure~\ref{fig:result:peaks}(a) we show the radii corresponding to the measured $Q_{\mathrm{g}}$ and the radii of the \qty{}{\kilo \parsec}-scale gas over-densities. Since on average $\sim \qty{2}{}$ gas peaks could be identified for each radial profiles, we choose the nearest one to the maximal torque force. In Figure~\ref{fig:result:peaks}(a) suggest that the radial offset between $Q_{\mathrm{g}}$ ($Q_{\mathrm{g,\, +gas}}$) and the gas over-density is minimal, especially when we focus on galaxies with strong torque forces (barred \textit{and} non-barred galaxies with $Q_{\mathrm{g}}\gtrsim \qty{0.05}{}$), suggesting gas over-density occurring close to $Q_{\mathrm{g}}$. The median radial offset from $Q_{\mathrm{g}}$ to the nearest gas over-densities is $\sim \qty{0.07}{}R_{25.5}$ ($\sim \qty{0.08}{}R_{25.5}$ for $Q_{\mathrm{g,\, +gas}}$). This is to be compared with the median complying to null hypothesis, which is $\sim \qty{0.16}{}R_{25.5}$, assuming that $Q_{\mathrm{g}}$ and the \qty{2}{} gas over-densities occur randomly within $R_{25.5}$. All galaxies with $Q_{\mathrm{g}}\gtrsim \qty{0.05}{}$ have radial offsets smaller than this median complying to null hypothesis. This suggest that the observed proximity between the maximal torque force and the \qty{}{\kilo \parsec}-scale gas over-densities does not seem to be a coincidence.

In comparison, Figure~\ref{fig:result:peaks}(b) suggests that the maximal $m=2$ surface density Fourier amplitude $\max \left\{ A_2\right\}$ radius and the (corresponding nearest) gas peak radius show a distribution that complies more to randomness. Data-points with stronger $\max \left\{ A_2\right\}$ do not seem to show smaller radial offsets. The median radial offset for $\sim \qty{0.12}{}R_{25.5}$, close to the expectation for no connection.

Our results are thus consistent with the scenario where sufficiently strong non-axisymmetric structures, especially when satisfying $Q_{\mathrm{g}}\gtrsim \qty{0.05}{}$, might directly induce \qty{}{\kilo \parsec}-scale neutral gas over-densities in their vicinity. As has been mentioned, a likely mechanism is the induced shock processes, which compress the gas along the leading edges of bar \cite[e.g.][]{1992MNRAS.259..345A}, as gas enters the spiral gravitational potential \cite[e.g.][]{2014MNRAS.440..208K}, or where gas orbits crowd, intersect or auto-sect \cite[]{2003ApJ...582..723R, 2004ApJ...600..595R, 2015MNRAS.454.3299R}. It is to be noted that, in addition to the shock-related processes, several other mechanisms have also been proposed to lead to small-scale gas over-densities, such as dynamical resonances \cite[]{1995ApJ...449..508P, 2000A&A...362..465R}, accumulation of inflowing gas where the centrifugal force balances gravity \cite[]{2012ApJ...747...60K} or where the shear reduces \cite[]{2015MNRAS.453..739K}. These mechanisms might provide explanation for gas over-densities that occur far from the maximal torque force radius.

\section{Discussion} \label{sec:discussion}

\subsection{The Advantage of $Q_{\mathrm{T}}$ in Characterizing the Driven Gas Inflow} \label{subsec:implication}

As has been discussed in Section~\ref{sec:intro}, non-axisymmetric structures (bars and spiral arms) share a uniformity in their role to interact with the neutral gas and drive the gas inward. Stronger disk non-axisymmetric structures could arguably perturb the gas motion and drive gas inflow in a more efficient manner \cite[e.g.][]{1992MNRAS.259..345A, 2012ApJ...758...14K, 2014MNRAS.440..208K}.

Our finding suggests that non-axisymmetric structures are able to affect \textit{neutral gas concentration on the disk-scale} (in the present work, the scale of $\qty{0.5}{} R_{25.5}$). This is roughly $\sim \qty{2}{}$ times the optical half-light radius for typical star-forming galaxies, and on average corresponding to a physical scale of $\sim \qty{13.5}{\kilo \parsec}$ in our sample. This complements previous works demonstrating enhanced central-\qty{}{\kilo \parsec} molecular gas content and star formation in barred galaxies \cite[e.g.,][]{1999ApJ...525..691S, 2005ApJ...632..217S, 2007PASJ...59..117K, 2012MNRAS.423.3486W,2017ApJ...838..105L, 2019MNRAS.484.5192C} or galaxies with strong spiral arms \cite[]{2022AaA...666A.175Y} that presumably leads to the growth of pseudo-bulges \cite[]{2004ARA&A..42..603K, 2005MNRAS.358.1477A, 2022A&A...661A..98Y}.

The efficiency of non-axisymmetric structures in driving this neutral gas concentration is seen to depend on \textit{their strength to perturb the gas motion}. The radially average torque force $\bar Q_{\mathrm{T}}$ seems to be more intrinsic in quantifying this strength, compared with the more conventional method based on the Fourier amplitudes of stellar mass surface density \cite[e.g.,][]{2022AaA...666A.175Y}. In this regard, we highlight that these two methods of characterizing the non-axisymmetric structures have different focuses and are arguably complementary. The more conventional characterization based on the stellar mass surface density measures the morphological strength of non-axisymmetric structures in an absolute sense, and is helpful in understanding the formation and temporal evolution of non-axisymmetric structures \cite[e.g.,][]{2002MNRAS.330...35A, 2003MNRAS.341.1179A, 2013MNRAS.429.1949A} dependent upon galaxy properties \cite[e.g.,][]{2016AAA...587A.160D, 2021ApJ...917...88Y}. In contrast, $\bar Q_{\mathrm{T}}$ quantifies not only the tangential force due to non-axisymmetric structures, but also considers factors such as stabilization due to bulge and dark matter halo \cite[]{2004MNRAS.355.1251L, 2016AAA...587A.160D} that counter their actions. The $\bar Q_{\mathrm{T}}$ parameter thus better highlights the dynamical strength of non-axisymmetric structures, which is more directly related to the gas motion and the gas redistribution \cite[]{2005A&A...441.1011G, 2013ApJ...771...59M, 2009ApJ...692.1623H, 2015MNRAS.451..936S}. In this sense, the findings of the present work further provides a physical aspect for how the gas inflow leading up to the observed inner gas concentration is happening. 

\subsection{The Non-Trivial Gas Inflow} \label{subsec:inflow}

\begin{figure}
    \centering
    \includegraphics[width=\columnwidth]{"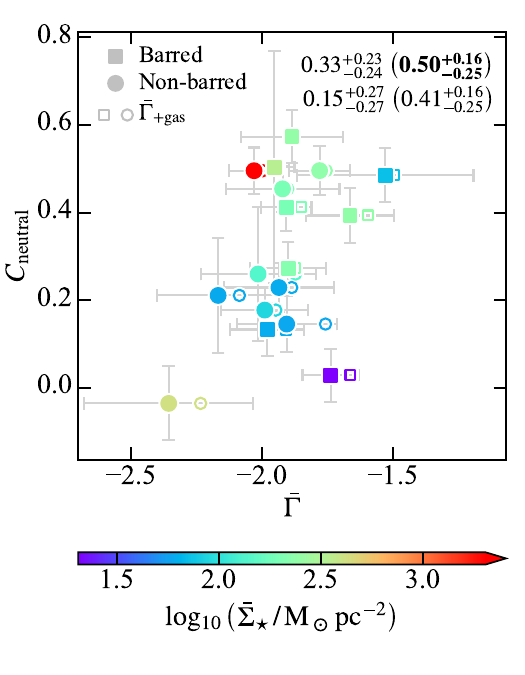"}
    \caption{The dynamical torque term $\bar{\Gamma}$ and the inner neutral gas concentration parameter $C_{\mathrm{neutral}}$. Symbols color, style, and annotations are the same to Figure~\ref{fig:result:im_c05}. A positive trend might be seen, albeit with significantly larger scatter, and only of marginal statistical significance at best.}
    \label{fig:result:tq_c}
\end{figure}

An interesting finding of the present work is that both the average torque forces and the maximal torque forces correlate with the inner neutral gas concentration parameter. It would thus be interesting to distinguish which of the two correlations is more statistically significant, thus likely more intrinsic. If the correlation between gas distribution and the average value is more intrinsic, then it is preferable to conjecture that non-axisymmetric structures likely work together or in relay to transport the gas across the disk \cite[e.g.,][]{2012MNRAS.423.3486W, 2012A&A...548A.126M, 2020ApJ...893...19W}. Otherwise, the strongest and the most dominant component likely dominates or controls over the processes. Indirect arguments from the literature exists. For example, \cite{2013ApJ...769..100S} have noted that in their simulations, without fuelling, e.g., from the spiral arms \cite[]{2014ApJ...782...64K}, the starbursts in the nuclear ring of barred galaxy dwindle quickly (often in $\lesssim \qty{0.1}{\giga \yr}$), which we note are inconsistent with observations demonstrating more persistent star formation enhancements \cite[$\gtrsim \qty{1}{\giga \yr}$ though possibly episodic,][]{2013A&A...551A..81V, 2017ApJ...838..105L, 2020MNRAS.499.1406L}. Observationally, \cite{2020ApJ...893...19W} have demonstrated that barred galaxies with central star formation enhancement/suppression preferentially have stronger/weaker spiral arms. These authors have proposed a possible scenario in which various disk morphological features have different pattern speeds $\Omega_ \mathrm{p}$ [as has been expected by theories such as \cite{2012A&A...548A.126M} and indeed been found in observations such as \cite{2009ApJS..182..559B}] such that they could each transport gas inward within the respective co-rotation resonance \cite[]{1987ApJ...318L..43T, 2004ARA&A..42..603K}, thus forming concentrated gas distribution more effectively. Sadly, a meticulous comparison between the average and the maximal torque forces is hindered by the small sample size of the present work, and we leave this as an open issue for future investigations.

While we highlight the use of the ``torque forces parameter'' $\bar Q_{\mathrm{T}}$ to characterize the non-axisymmetric structures, we do not intend to claim that the dynamical torque is the only or the dominant mechanism in the inward transportation of the neutral gas. In Figure~\ref{fig:result:tq_c}, we show the dynamical torque term defined as
\begin{equation}
    \bar{\Gamma}=\mathrm{average}\left\{ \frac{r}{R_{25.5}}\cdot \frac{f_{\mathrm{T}}}{f_r} \right\} ,\qquad 3\gamma _{\mathrm{PSF}}^{\prime}<r<R_{25.5},
\end{equation}
mimicking the textbook definition of gravitational torque, $\boldsymbol{\Gamma}=\mathbf{r}\times \mathbf{f}$, and the inner neutral gas concentration $C_{\mathrm{neutral}}$. A correlation between $\bar \Gamma$ and $C_{\mathrm{neutral}}$ is expected if we envision a situation in which the gas hardly dissipates, being collected and funnelled inward only passively with the dynamical torque being the sole channel for the gas to lose angular momentum. However, Figure~\ref{fig:result:tq_c} suggests that this does not seem to be the case, since larger scatter compared to the correlation between $\bar Q_{\mathrm{T}}$ and $C_{\mathrm{neutral}}$ (Figure~\ref{fig:result:im_c05}) seems to be seen, with the correlation appearing to be only marginally statistically signifiant at best, seen from Spearman's statistics. In this regard, \cite{2014ApJ...782...64K} have demonstrated in their simulations that, in principle, three mechanisms might be at work in transporting the gas inward, i.e., (\romannumeral1) gravitational torque due to stellar non-axisymmetric structures, (\romannumeral2) due to the gas self-gravity, and (\romannumeral3) dissipative processes, e.g., shocks. These authors have estimated that these three mechanisms each contribute $\sim \qty{40}{\percent}$, $\sim \qty{10}{\percent}$ and $\sim \qty{50}{\percent}$, respectively, to the mass inflow rate averaged between the inner Lindblad and the co-rotation resonances. In this regard, gas dissipation is arguably a more important mechanism \cite[e.g.,][]{1992MNRAS.259..345A,2012ApJ...758...14K,2014ApJ...789...68K}, and our $\bar Q_{\mathrm{T}}$ measures more intrinsically the efficiency of non-axisymmetric structures in perturbing the gas motion, and inducing these dissipative processes. Future observations directly comparing gas kinematics \cite[e.g.][]{2016MNRAS.457.2642S, 2021ApJ...923..220D, 2022ApJ...939...40L} and gravitational force and torque estimation \cite[e.g.][]{2005A&A...441.1011G, 2009ApJ...692.1623H, 2016A&A...588A..33Q, 2024arXiv241001721D} might be needed to constrain the contribution of each mechanisms in the real Universe.

To summarize, we emphasize that, these complexities of gas inflow driven by non-axisymmetric structures do emerge only when the physically motivated radially average torque forces $\bar Q_{\mathrm{T}}$ parameter is used to quantify the effects related to the non-axisymmetric structures.

\subsection{Other Processes Affecting the Neutral Gas Distribution} \label{subsec:other}

Apart from the non-axisymmetric structures, several other mechanisms or processes might also affect the neutral gas distribution, which we briefly discuss here for the purpose of completeness. \textit{Viscosity} \cite[]{1974MNRAS.168..603L, 1981MNRAS.194..967B} affects gas motion significantly at a $\sim$ \text{\qtyrange[range-phrase=--, range-units=bracket]{100}{200}{\parsec}}-scale, yet it should be rather inefficient at \qty{}{\kilo \parsec}-scale \cite[for an estimation of viscous timescale as a function of galactocentric radius, see][]{2005A&A...441.1011G}. \textit{Galactic fountain} \cite[]{2012MNRAS.419..771B, 2019MNRAS.490.4786G} might enhance the accretion of gas with lower angular momentum, raising the concentration in roughly a few \qty{}{\giga \yr} \cite[]{2014ApJ...796..110E}, yet it should be universal and thus is irrelevant to the observed correlations between gas distribution and the torque parameter.

\textit{Environmental effects} are arguably more intricate mechanism, which affect both the neutral gas distribution \cite[e.g.,][]{2021ApJ...915...70W, 2022ApJ...927...66W, 2023ApJ...956..148L, 2019MNRAS.484.5192C}, and induce or enhance the non-axisymmetric structures \cite[]{1991ApJ...370L..65B, 2014MNRAS.445.1339L, 2019MNRAS.483.2721P} which might be long-lived \cite[up to $\sim \qty{10}{\giga \yr}$,][]{2014MNRAS.445.1339L, 2016ApJ...826..227L, 2019MNRAS.483.2721P}. For a vivid example, while dynamical modelling has suggested that the grand-design spiral arms of \messier{51a} (\ngc{5194}) is of tidal origin \cite[]{2010MNRAS.403..625D}, \cite{2016A&A...588A..33Q} have suggested observational evidence for gas inflow purely due to the spiral gravitational torque. Disentangling these effects are beyond the scope of the present work, which makes use of a small sample.

\subsection{Implications on Galaxy Evolution Driven by the Non-Axisymmetric Structures} \label{subsec:imply}

Recent investigations with \acro{jwst} and \acro{alma} have found that galactic disks \cite[]{2022ApJ...938L...2F, 2023ApJ...942L...1W, 2023ApJ...946L..15K, 2023ApJ...948L..18N, 2023ApJ...956..147H, 2024ApJ...960..104S, 2024ApJ...960L..10L}, some of which are dynamically cold \cite[]{2021A&A...647A.194F, 2024arXiv240416963X, 2024MNRAS.tmp.2235R}, as well as bars and spiral arms therein \cite[]{2023ApJ...942L..42R, 2023ApJ...945L..10G, 2023ApJ...958L..26H, 2024MNRAS.530.1984L, 2024ApJ...968L..15K}, seem to develop early in the cosmic time. These exciting findings hint at a possible revision to the secularly versus merger-built bulge-like structures in the Universe. Since the \textit{Hubble Space Telescope} (\textit{HST}) campaigns on several deep galaxy fields, it has been conventional conduct to count fraction of galaxies hosting bars or spiral arms as a function of redshift, to evaluate the formation, growth and weakening of these structures, as well as their impact on secular evolution in the galaxy lifetime \cite[]{2008ApJ...675.1141S, 2014MNRAS.438.2882M, 2014MNRAS.445.3466S, 2014ApJ...781...11E}. Our results suggest that specific studies on the impact of single, particular morphological features on the build-up of galaxy central stellar mass concentration could be unified within the same physical framework of non-axisymmetric structures. Such a unified view is important considering the fact that the non-axisymmetric structures are more complex than just bars and spiral arms, and that these structures might not be highly independent of each other. For example, $m=1$ non-axisymmetric structures induced by tidal interactions or gas accretion are more common at high-$z$ \cite[]{2022A&A...666A..44K, 2024A&A...688A..53L}; the kpc-scale clumps produced in turbulent disks at high-$z$ could be much more signifiant non-axisymmetric structures than those seen at low-$z$ in the appearance of bars and spiral arms \cite[]{2008ApJ...688...67E, 2024ApJ...960...25K}; bars and spiral arms are often triggered by the same perturbation that produces $m=1$ structures \cite[e.g., tidal interactions,][]{2014MNRAS.445.1339L}; bars and spiral arms could be coupled with one another \cite[]{2010ApJ...722..112M, 2010ApJ...715L..56S}, and bars could dissolve into lenses which might still be significantly non-asymmetric \cite[]{2004ARA&A..42..603K}. 

These relevant effects should be re-evaluated not only in terms of observations but also in terms of simulations. The predictions from the state-of-art cosmological hydrodynamical simulations still deviate significantly from the observational trend in the size-mass relation of relatively passive, thus more bulge-dominated, galaxies, while the disks produced in these simulations are often too hot and too thick \cite[see][and references therein]{2023ARA&A..61..473C}. A cold dynamical state of the stellar disks is necessary for the formation of non-axisymmetric structures \cite[]{2012ApJ...757...60K, 2012ApJ...758..136S, 2013MNRAS.429.1949A, 2019ApJ...872....5S}, while the gas inflow driven by these structures is important in fuelling inner disk star formation and pseudo-bulge formation \cite[]{2004ARA&A..42..603K}. It is therefore reasonable to speculate that, future simulations must first reproduce realistic non-axisymmetric structures, including bars and spiral arms, simultaneously, before they could physically solve the size-mass relation problem.

The gravitational tangential forces evaluated in the present work provide a physically motivated simple tool to avoid biased focuses on a single particular morphological feature thus to link the related processes and aspects of non-axisymmetric and axisymmetric structures formation, as well as to bridge the gap between simulations and observations. Such quantification on the observational side would be feasible for large samples with the up-coming wide-field, high-quality imaging surveys, including the High Latitude Wide Area Survey \cite[\acro{hlwas};][]{2015arXiv150303757S, 2019arXiv190205569A} with the \textit{Nancy Grace Roman Space Telescope}, the \textit{Euclid} Wide Survey \cite[\textsc{ews};][]{2022A&A...662A.112E} with the \textit{Euclid} Space Telescope, the Legacy Survey of Space and Time \cite[\acro{lsst};][]{2019ApJ...873..111I} with the Vera C. Rubin Observatory and the Chinese Space Station Optical Survey \cite[\acro{css-os};][]{2011SSPMA..41.1441Z, 2018..............Z, 2018MNRAS.480.2178C, 2019ApJ...883..203G} with the Chinese Space Station Telescope (\acro{csst}).

\subsection{Limitations and Caveats} \label{subsec:caveat}

Before bringing our discussion to an end, it is important to reiterate that the results of the present work are based on a sample limited in size. The limitation stems from the need for a set of uniform and high-resolution neutral gas imaging data capable of resolving the disk. We acknowledge that the sample size has inevitably limited the statistical significance of our work. The goal of this work is thus to establish the existence of potential correlations between observables, and future observations of larger sample sizes are needed to further investigate the statistical significance of these correlations.

Due to the small sample size, we are not able to take into account the potential impact of environmental effects on galaxy evolution (Section~\ref{subsec:other}). Future studies that probe a significant cosmic volume will further deepen our understanding of the complicated interplay between different processes in galaxy evolution.

\section{Summary and Conclusion} \label{sec:summary}

In the present work, we demonstrate the existence of potential correlation between the \textit{strength} of non-axisymmetric structures, quantified by the exerted torque forces, and the distribution of the \textit{total neutral} (i.e., atomic and molecular) gas in disk galaxies. With \qty{}{17} galaxies selected from the \things sample, we estimate the torque field $Q_{\mathrm{T}}$ using the \qty{3.6}{\micro \metre} images (Section~\ref{sec:data}). Recognizing the importance of properly calculating the forces due to the bulge component and the dark matter halo, we improve the bulge stretching correction with optimal decomposition strategy and take into consideration the stabilizing effects of the dark matter halo (Section~\ref{sec:method}). The evaluated $Q_{\mathrm{T}}$, which we believe is a physically motivated estimation for the strength of the non-axisymmetric structures, is then used to examine the correlations with disk-scale neutral gas distribution (Section~\ref{sec:result}).

We summarize our major results as follows.
\begin{enumerate*}
    \item[\romannumeral1.] Galaxies with stronger average torque forces $\bar Q_{\mathrm{T}}$ due to non-axisymmetric structures tend to have more concentrated disk-scale neutral gas distribution within the inner disk (quantified by $C_{\mathrm{neutral}}$; Figure~\ref{fig:result:im_c05}). The enhanced inner neutral gas concentration seems unlikely to be explained fully by the enhanced molecular gas content in the central-\qty{}{\kilo \parsec} regions (quantified by $C_{\mathrm{CO}}$) as is known in the literature [Figure~\ref{fig:result:corrs}(a)].
    \item[\romannumeral2.] Controlling for either the disk gas fraction $f_{\mathrm{neutral}}\left( R_{25.5}\right)$ or the average stellar mass surface density $\bar \Sigma _{\star}$ does not significantly affect the correlation between $\bar Q_{\mathrm{T}}$ and $C_{\mathrm{neutral}}$, suggesting that the complex gas regulation cannot fully explain the observed trend (Section~\ref{subsec:partial}).
    \item[\romannumeral3.] We do not observe significant correlation of $2+\log _{10}s_{\mathrm{disk}}$ with $C_{\mathrm{neutral}} $ or $C_{\mathrm{CO}}$, likely suggesting that torque forces are a more intrinsic measurable quantifying the dynamical strength of disk non-axisymmetric structures [Figure~\ref{fig:result:corrs}(b, c)].
    \item[\romannumeral4.] Inner neutral gas concentration $C_{\mathrm{neutral}}$ correlates with the maximal torque forces $Q_{\mathrm{g}}$ as well [Figure~\ref{fig:result:corrs}(d)], raising the problem of which of the two correlations ($\bar Q_{\mathrm{T}}$--$C_{\mathrm{neutral}}$ or $Q_{\mathrm{g}}$--$C_{\mathrm{neutral}}$) might be more intrinsic.
    \item[\romannumeral5.] Provided that the torque forces are locally sufficiently strong ($Q_\mathrm{g}\gtrsim 0.05$), the \qty{}{\kilo \parsec}-scale neutral gas over-densities preferentially occur close to the maximal torque forces radius, suggesting connection. When the gravitational perturbation is less strong ($Q_\mathrm{g}\lesssim 0.05$), there does not seem to be such a connection (Figure~\ref{fig:result:peaks}).
\end{enumerate*}

Our results are consistent with the secular evolution scenario in which the non-axisymmetric structures perturb the motion of gas, trigger shock processes, and funnel the gas inward within the galactic disk. Non-axisymmetric structures share a uniform role in this process, with the effectiveness dependent upon the strength to dynamically perturb the gas motion, quantifiable by the torque force parameter $\bar Q_{\mathrm{T}}$  (Section~\ref{sec:discussion}).

The limitations of the present work mainly concern the small sample size (Section~\ref{subsec:caveat}). Future studies extending the methodology to larger samples would be needed to re-examine the generality of conclusions in the present work, as well as to revisit the role of non-axisymmetric structures in galaxy secular evolution.


\begin{acknowledgments}

We are extremely grateful to the anonymous referee for constructive comments and practical suggestions that greatly improved this work. LCH was supported by the National Science Foundation of China (11991052, 12011540375, 12233001), the National Key R\&D Program of China (2022YFF0503401), and the China Manned Space Project (CMS-CSST-2021-A04, CMS-CSST-2021-A06). This work made use of \things, ``The \ion{H}{1} Nearby Galaxy Survey'' \cite[]{2008AJ....136.2563W}. This work made use of \heracles, ``The HERA CO-Line Extragalactic Survey'' \cite[]{2009AJ....137.4670L}. This research has made use of the NASA/IPAC Extragalactic Database (NED), which is funded by the National Aeronautics and Space Administration and operated by the California Institute of Technology. We acknowledge the usage of the HyperLeda database \cite[\url{http://leda.univ-lyon1.fr};][]{2014A&A...570A..13M}. This work made use of \textsc{Astropy}:\footnote{http://www.astropy.org} a community-developed core Python package and an ecosystem of tools and resources for astronomy \cite[]{2013A&A...558A..33A, 2018AJ....156..123A, 2022ApJ...935..167A}. This research made use of \textsc{Photutils}, an \textsc{Astropy} package for detection and photometry of astronomical sources \cite[]{2022zndo...6825092B}.

\end{acknowledgments}

%

\facilities{\textit{The Spitzer Space Telescope}, VLA (NRAO), BIMA, The 30-m Telescope (IRAM), The \textit{Wide-field Infrared Survey Explorer}, The \textit{Galaxy Evolution Explorer}}


\software{
    \textsc{Astropy} \cite[]{2013A&A...558A..33A, 2018AJ....156..123A, 2022ApJ...935..167A},
    \textsc{Imfit} \cite[]{2014ascl.soft08001E, 2015ApJ...799..226E},
    \textsc{matplotlib} \cite[]{2007CSE.....9...90H},
    \textsc{NumPy} \cite[]{2020Natur.585..357C},
    \textsc{Photutils} \cite[]{2022zndo...6825092B},
    \textsc{Pingouin} \cite[]{2018JOSS....3.1026V},
    \textsc{ProFound} \cite[]{2018ascl.soft04006R, 2018MNRAS.476.3137R},
    \textsc{PyRAF} \cite[]{2001ASPC..238...59D, 2012ascl.soft07011S},
    \textsc{reproject} \cite[]{2018zndo...1162674R},
    \textsc{SciPy} \cite[]{2020NatMe..17..261V},
    \textsc{source-extractor} \cite[]{1996A&AS..117..393B}
}



\newpage
\appendix

\section{Uncertainty in Torque Force Evaluation} \label{sec:uncertainties}


\begin{deluxetable}{ccccc}
    \tabletypesize{\footnotesize}
    \tablewidth{0pt}
    \tablecaption{Uncertainty Analysis for $\bar Q_{\mathrm{T}}$}
    \label{table:uncertainty}
    \tablehead{
        \colhead{Index} &
        \colhead{Item} &
        \multicolumn{2}{c}{Uncertainty} &
        \colhead{Note} \\
        \colhead{} &
        \colhead{} &
        \colhead{Absolute} &
        \colhead{Relative} &
        \colhead{}
    }
    \colnumbers
    \startdata
    1 & Bulge correction omitted & $\sim \qty{0.003}{}$ & $\sim \qty{13}{\percent}$ & * \\
    2 & Bulge shape & $\sim \qty{0.0003}{}$ & $\sim \qty{1.2}{\percent}$ & * \\
    3 & Decomposition statistical error & $\sim \qty{0.0003}{}$ & $\sim \qty{0.9}{\percent}$ & * \\
    4 & Disk scale height & $\sim \qty{0.0017}{}$ & $\sim \qty{5}{\percent}$ &  \\
    5 & Integration boundary & $\sim \qty{0.000006}{}$ & $\sim \qty{0.025}{\percent}$ &  \\
    6 & Set inner harmonics to \qty{0}{} & $\sim \qty{0.00030}{}$ & $\sim \qty{0.9}{\percent}$ &  \\
    7 & Higher even-order harmonics & $\sim \qty{0.0013}{}$ & $\sim \qty{4}{\percent}$ & $\dagger$ \\
    8 & Higher odd-order harmonics & $\sim \qty{0.0022}{}$ & $\sim \qty{7}{\percent}$ & $\ddagger$ \\
    9 & Inclination $i$ & $\sim \qty{0.004}{}$ & $\sim \qty{18}{\percent}$ &  \\
    10 & Position angle $\phi _\mathrm{PA}$ & $\sim \qty{0.004}{}$ & $\sim \qty{15}{\percent}$ &  \\
    11 & Center & $\sim \qty{0.0020}{}$ & $\sim \qty{6}{\percent}$ &  \\
    \hline
    12 & All & $\sim \qty{0.013}{}$ & $\sim \qty{48}{\percent}$ & $\S$ \\
    \hline
    13 & Dust emission correction & $\lesssim \qty{0.010}{}$ & $\lesssim \qty{39}{\percent}$ & $\parallel$ \\
    14 & Neutral gas self-gravity & $\sim \qty{0.0027}{}$ & $\sim \qty{10}{\percent}$ &  \\
    \enddata
    \tablecomments{Uncertainty estimation for the average torque force $\bar Q_{\mathrm{T}}$. (1) Index. (2) Item considered. (3) Absolute uncertainty propagate to $\bar Q_{\mathrm{T}}$. (4) Relative uncertainty propagate to $\bar Q_{\mathrm{T}}$. (4) Note. \\
        \\
        *. Bulge-less galaxies are omitted from consideration. \\
        $\dagger$. Azimuthal Fourier harmonics $m=10,\, 12,\, 14,\, \cdots,\, 20$ are included in evaluation in addition to the default $m=0,\, 1,\, 2,\, 3,\, 4,\, 6,\, 8$, to evaluate the contribution of higher even-order harmonics. \\
        $\ddagger$. Azimuthal Fourier harmonics $m=5,\, 7,\, 9$ are included in evaluation in addition to the default to evaluate the contribution of higher odd-order harmonics. \\
        $\S$. Total uncertainty is estimated by summing all terms listed above in quadrature. \\
        $\parallel$. Only galaxies with the ICA stellar mass (s1) maps available from \cite{2015ApJS..219....5Q} are considered. Some galaxies with low dust emission are thus excluded by this selection.
    }
\end{deluxetable}


Apparently, evaluation of the torque field outlined in Section~\ref{sec:method} depends upon the multi-component decomposition (Section~\ref{subsec:bulge}), selection of the Fourier components, and the adopted disk scale height, inclination, and position angle. We run tests evaluating the torque field with alternation in inputs to estimate the uncertainties associated with the torque field evaluation for each of our sample galaxies, propagated from aforementioned sources. The results are summarized in Table~\ref{table:uncertainty}, and as follows.

Overall, our estimation suggests that the most significant contributors of uncertainties in $\bar Q_{\mathrm{T}}$ include (\romannumeral1) disk inclination (contributing to an absolute uncertainty in $\bar Q_{\mathrm{T}}$ of order $\sim \qty{0.004}{}$, or a relative uncertainty of $\sim \qty{18}{\percent}$, corresponding to errors reported in Table~\ref{table:sample}), (\romannumeral2) disk position angle ($\sim \qty{0.004}{}$, or $\sim \qty{15}{\percent}$, corresponding to errors reported in Table~\ref{table:sample}), (\romannumeral3) bulge stretching correction ($\sim \qty{0.003}{}$, or $\sim \qty{13}{\percent}$, if completely omitted). It could be seen that the uncertainties associated with the bulge stretching correction increases with decreasing radius, while those propagated from inclination and position angle are significant throughout all radial range. The result agrees with the literature suggesting that the disk inclination and position angle could be a major source of uncertainty, in that any errors in the disk orientation would lead to artificially distorted disk in de-projection \cite[]{2009ApJ...692.1623H, 2016A&A...588A..33Q}. The large uncertainty associated with the bulge projection underscores the importance of bulge stretching correction \cite[]{2004MNRAS.355.1251L}, and necessitates the employment of optimal multi-component decomposition to further prevent the potentially dramatic error in the decomposition itself \cite[]{2017ApJ...845..114G, 2015APJS..219....4S}.

Within the inner disk, the uncertainties introduced by (\romannumeral4) the central coordinate (overall, $\sim \qty{0.0020}{}$, or $\sim \qty{6}{\percent}$, assuming precise to $\gamma _{\mathrm{PSF}}$) and (\romannumeral5) the assumed disk scale height \cite[overall, $\sim \qty{0.0017}{}$, or $\sim \qty{5}{\percent}$, corresponding to scatter reported in][]{1998MNRAS.299..595D} could also be significant sources of uncertainties. Though they contribute minimally to the overall uncertainty of $\bar Q_{\mathrm{T}}$ (averaged within $R_{25.5}$), their contribution to the local estimation $Q_{\mathrm{T}}$ could be greater than $\gtrsim \qty{0.003}{}$, or $\gtrsim \qty{10}{\percent}$ within $\sim$ \qtyrange[range-phrase=--, range-units=bracket]{0.1}{0.2}{}$R_{25.5}$, the typical bar radius \cite[]{2016AAA...587A.160D}. Investigation focussing on the inner disk or the inner non-axisymmetric structures (e.g., bars) might therefore find it necessary to consider these uncertainties \cite[e.g., as in][]{2002MNRAS.337.1118L, 2004MNRAS.355.1251L, 2016AAA...587A.160D}.

Some uncertainties arise from the assumption of the bulge shape and the harmonics included in the integration, yet they are of limited magnitude and should be sound to omit in simplified analyses. Assuming symmetric yet disk-like bulge \cite[]{2005MNRAS.358.1477A} leads to differences of order $\sim \qty{0.0003}{}$, or $\sim \qty{1.2}{\percent}$ compared to the default assumption adopted in the present work (i.e., symmetric and spherical), with the uncertainty increases with decreasing radius. Compared with the uncertainty that would have been introduced by omitting the bulge stretching correction, the uncertainty associated with the bulge shape would be practically negligible. Including higher even-order and odd-order harmonics leads to differences of order $\sim \qty{0.0013}{}$ and order $\sim \qty{0.0022}{}$, or $\sim \qty{4}{\percent}$ and $\sim \qty{7}{\percent}$, respectively, across all the radial range. Despite so, as has been argued in Section~\ref{sec:method}, including the higher order harmonics risks biases by sharp features due to, e.g., low signal-to-noise ratio or dust emission \cite[]{2002MNRAS.337.1118L, 2016A&A...596A..84D}, which would be undesirable for robust estimation.

In comparison, statistical uncertainties associated with the optimal multi-component decomposition \cite[estimated following the method outlined in][]{2015APJS..219....4S}\footnote{In the present work, the differences between the cryogenic and the post-cryogenic PSF is additionally considered together with the uncertainties associated with the sigma-map, background subtraction and the asymmetry of PSF considered in \cite{2015APJS..219....4S}.} propagate minimally to the uncertainty in $\bar Q_{\mathrm{T}}$. Setting non-axisymmetric harmonics inside the bulge-dominated region to \qty{0}{} also introduces negligible difference suggesting little impact of nuclear axisymmetric structures \cite[]{1989Natur.338...45S, 2004ApJ...617L.115E} on the global dynamics, justifying the omission of nuclear morphological features in multi-component decomposition of a large sample \cite[]{2019ApJS..244...34G}. Neglecting the inner non-axisymmetric harmonics would also suppress the sometimes large residual resulted from the multi-component decomposition. In terms of convergence, an outer integration boundary beyond $\gtrsim R_{25.5}$ is seen to be largely sound if one wishes to evaluate the torque field within this radius.

By summation in quadrature, one could estimate a total uncertainty typically of level $\sim \qty{0.013}{}$, or $\sim \qty{48}{\percent}$, in the evaluated $\bar Q_{\mathrm{T}}$. See also \cite{2002MNRAS.337.1118L} and \cite{2016AAA...587A.160D} for further arguments on how the further detailed assumptions on the disk vertical structures (the vertical profile and the radial gradient of scale-height) influences the torque field evaluation in an practically negligible manner.

There are also uncertainties that are difficult to estimate strictly in the scope of the present work, including (\romannumeral1) the mass-to-light ratio $\Upsilon _{\qty{3.6}{\micro \meter}}$, whose scatter is typically $\sim \qty{30}{\percent}$ \cite[]{2012AJ....143..139E}, (\romannumeral2) the scatter associated with the gravitational acceleration relation \cite[]{2016ApJ...827L..19L, 2017ApJ...836..152L}. \cite{2017ApJ...836..152L} which is typically \qtyrange[range-phrase=--, range-units=bracket]{0.13}{0.15}{\dex}, respectively, and eventually (\romannumeral3) the photometric calibration of IRAC imaging data and the aperture correction factors, which should each be accurate to $\sim \qty{3}{\percent}$ and $\sim \qty{10}{\percent}$,\footnote{See section~4.3 and section~8 of \cite{2021..............I}.} respectively. One would therefore expect that the actual uncertainty associated with the torque parameter $Q_{\mathrm{T}}$ to be slightly larger than that estimated in Table~\ref{table:uncertainty}.

\subsection{Dust Emission} \label{subsec:s1}

\begin{figure}
    \centering
    \includegraphics[width=\columnwidth]{"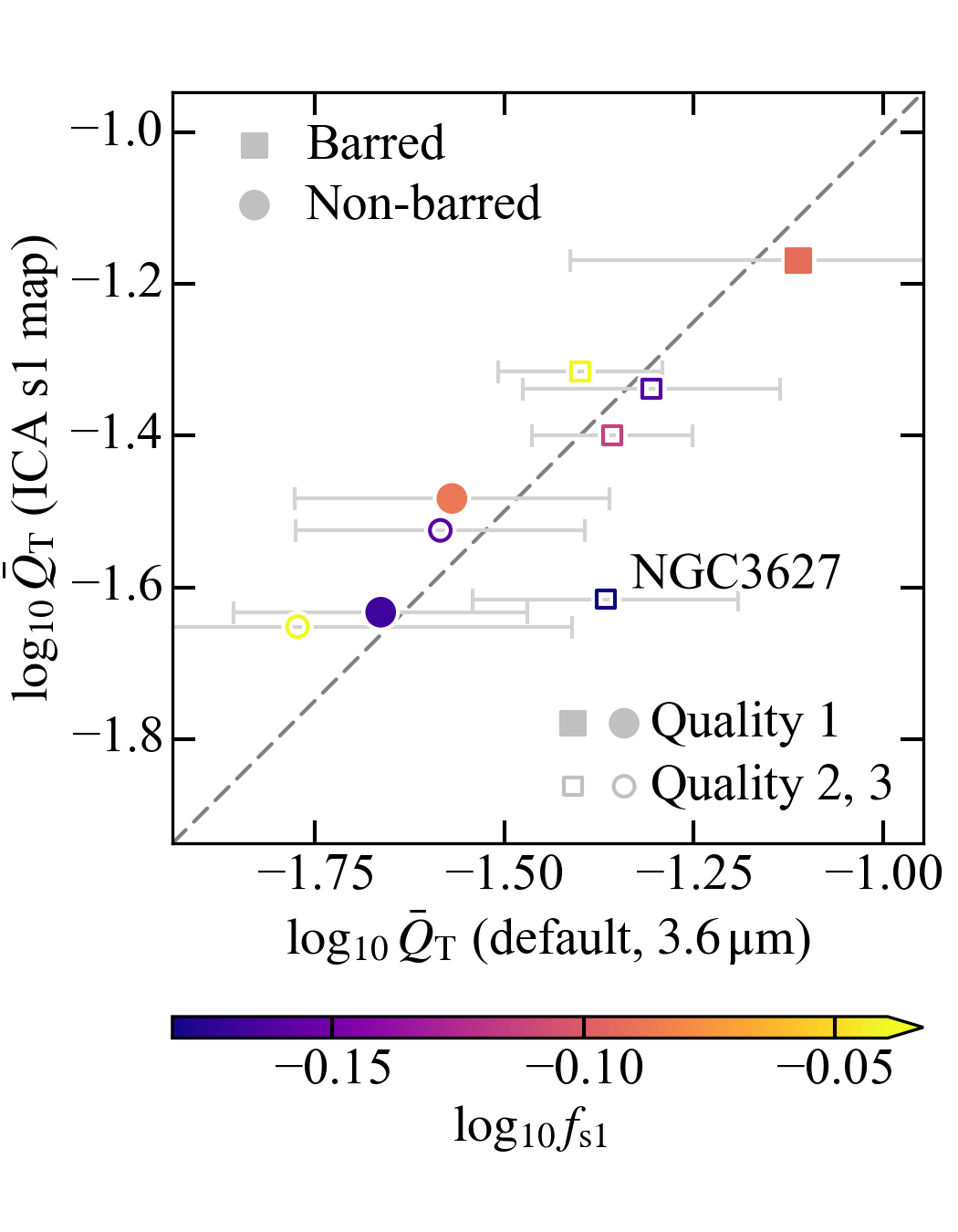"}
    \caption{Comparison between the default $Q_{\mathrm{T}}$ evaluation (based on \qty{3.6}{\micro \meter}), and that based on the ICA stellar mass (s1) maps derived by \cite{2015ApJS..219....5Q}, for \qty{9} sample galaxies. Symbols are color-coded according to s1-to-total flux ratio $f_{\mathrm{s1}}$, and are distinguished for barred (square) and non-barred (circle) galaxies. Filled symbols show cases with the most reliable ICA results, and open symbols show those that have been considered less reliable \cite[quality flag of $1$ or $2,\, 3$, respectively,][]{2015ApJS..219....5Q}. The dashed line represents $y=x$. General agreement is seen.}
    \label{fig:comp_ica}
\end{figure}

Mid-infrared imaging data at \qty{3.6}{\micro \meter} capture emission due to hot dust, PAH, as well as intermediate-age stars in addition to the light from the underlying old stellar populations \cite[]{2012ApJ...744...17M}. We test whether this former emission component could significantly affect the evaluation of the torque force $\bar Q_{\mathrm{T}}$. \cite{2015ApJS..219....5Q} have provided an estimation of the ratio between these two emission components based on the independent-component analysis (ICA) technique with the \sfourg \qtyrange[range-phrase={ / }]{3.6}{4.5}{\micro \meter} images \cite[]{2012ApJ...744...17M, 2015ApJS..219....5Q}. The derived ICA s1 (stellar mass) maps from \cite{2015ApJS..219....5Q} are available for \qty{10}{} out of our \qty{17}{} sample galaxies, which excludes \ngc{3031}, \ngc{5055} ($\left[ \qty{3.6}{\micro \meter}\right] -\left[ \qty{4.5}{\micro \meter}\right]$ color suggests dominance of old stellar population emission), \ngc{5457} (low signal-to-noise ratio), \ngc{925}, \ngc{2403}, \ngc{3621} and \ngc{6946} \cite[not included in the \sfourg sample,][]{2010PASP..122.1397S}. We do not consider \ngc{5194} in the test either, for the presence of \ngc{5195} in the field of view. These galaxies are thus omitted from the test.

Figure~\ref{fig:comp_ica} compares the average torque force $\bar Q_{\mathrm{T}}$ evaluated from the \qty{3.6}{\micro \meter} images and from the ICA s1 (stellar mass) maps derived by \cite{2015ApJS..219....5Q} that effectively trace the underlying old stellar populations. Note that we have not performed multi-component decomposition on the ICA s1 (stellar mass) map, since we consider it difficult to evaluate the corresponding pixel-wise uncertainty strictly. We thus expect some overestimation due to the bulge projection, which is indeed seen in Figure~\ref{fig:comp_ica} in which data-points tend to lie above the diagonal line. Beyond the bulge-dominated region, slight reduction is sometimes seen with the $Q_{\mathrm{T}}\left( r\right)$ profiles evaluated with the ICA s1 (stellar mass) maps, yet they generally agrees with those evaluated with the \qty{3.6}{\micro \meter} images within uncertainty. The considerably large underestimation seen for \ngc{3627} might be attributed to over-subtraction of the ICA technique preferentially along spiral arms, which, in the most extreme case, leads to zero flux along the prominent spiral arms. \cite{2015ApJS..219....5Q} have also considered that the ICA result for this galaxy is less reliable (by assigning a quality flag 2).

For the \qty{9}{} sample galaxies shown in Figure~\ref{fig:comp_ica}, we estimate the differences between $\bar Q_{\mathrm{T}}$ evaluated with the \qty{3.6}{\micro \meter} images and with the ICA s1 (stellar mass) maps to be of order $\lesssim \qty{0.010}{}$, or $\lesssim \qty{39}{\percent}$. Note that we regard this estimation as an upper limit here, as an sub-sample of galaxies overlapped with that of \cite{2015ApJS..219....5Q} is considered, and some galaxies with low dust emission are excluded by this selection. It thus seems that the dust emission does affect the evaluation of the torque field, but only in a limited manner compared to the uncertainties of other sources \cite[see also comparisons in][]{2016AAA...587A.160D}.

\subsection{Neutral Gas Self-Gravity} \label{subsec:ng}

\begin{figure}
    \centering
    \includegraphics[width=\columnwidth]{"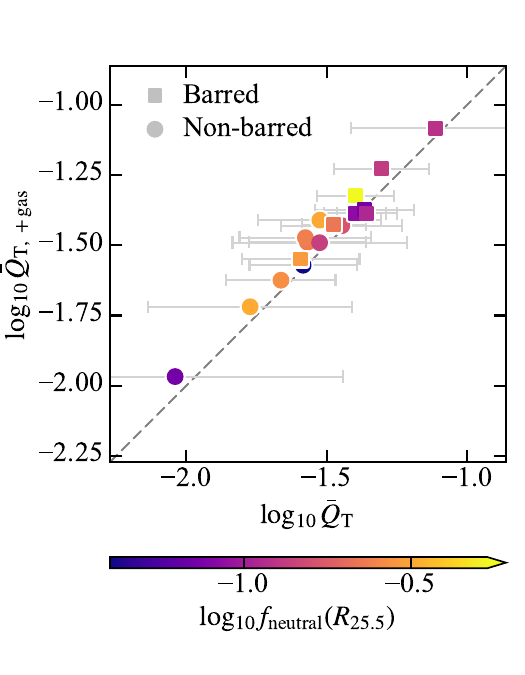"}
    \caption{Comparison between the default $\bar Q_{\mathrm{T}}$ evaluation (torque forces solely due to stars), and $\bar Q_{\mathrm{T,\, +gas}}$ that involves the gas self-gravity. Symbols are color-coded according to the neutral gas fraction $f_{\mathrm{neutral}}\left( R_{25.5}\right)$, and are distinguished for barred (square) and non-barred (circle) galaxies. The dashed line represents $y=x$. An increase dependent upon gas fraction is seen when gas self-gravity is considered, though of limited amplitude.}
    \label{fig:comp_neutral}
\end{figure}

We also evaluate the torque force field associated with the distribution of the neutral gas to understand its contribution. Neutral gas in disk could display rich structures due to various mechanisms including fragmentation, shocks, feedback, and hydrodynamical instability \cite[e.g.,][]{1992MNRAS.259..345A, 2014MNRAS.440..208K, 2014ApJ...789...68K, 2021ApJ...913..113M, 2023ApJ...944L..18M, 2023ApJ...944L..22B}, some of which \cite[e.g., feedback-driven shells, bubbles, and holes, stretched and elogated by shearing motion;][]{2009ApJ...704.1538W, 2019ApJ...872....5S, 2023ApJ...944L..24W, 2023ApJ...944L..22B} are considered to be transient compared to the disk dynamical timescale. We thus choose to convolve the gas distribution to a physical resolution of $\sim \qty{1.5}{\kilo \parsec}$, and omit the gas structures under this scale which we believe to be less coherent in affecting the motion of the gas. Even so, these feedback-driven features might still contribute to the non-axisymmetric part of the neutral gas self-gravity forces, since they could feature sizes up to $\sim \qty{1}{\kilo \parsec}$ \cite[]{2023ApJ...944L..22B}. We assume an infinitesimally thin gaseous disk, and omit the gas beyond the warp onset radius for \ngc{4736}, \ngc{5055}, and \ngc{7793} in this evaluation. We denote the average torque forces evaluated with the neutral gas self-gravity combined as $\bar Q_{\mathrm{T,\, +gas}}$.

From Figure~\ref{fig:comp_neutral}, one sees that inclusion of the gas self-gravity leads to an increase in the evaluated average torque forces (i.e., $\bar Q_{\mathrm{T,\, +gas}}>\bar Q_{\mathrm{T}}$ in general). Galaxies with more notable increases are seen to be comparatively richer in gas. Yet, the increase is in general limited in amplitude, with a typical difference of $\sim \qty{0.0027}{}$, or $\sim \qty{10}{\percent}$. This suggests that the neutral gas self-gravity contributes but to a limited extent to the overall torque force field. Furthermore, it seems that the non-axisymmetric part of the self-gravity forces is less closely related to the disk-scale neutral gas concentration (Section~\ref{subsec:average}).

\subsection{Optimal Multi-Component Decomposition} \label{subsec:comparison_decompose}

\begin{figure}
    \centering
    \includegraphics[width=0.95\columnwidth]{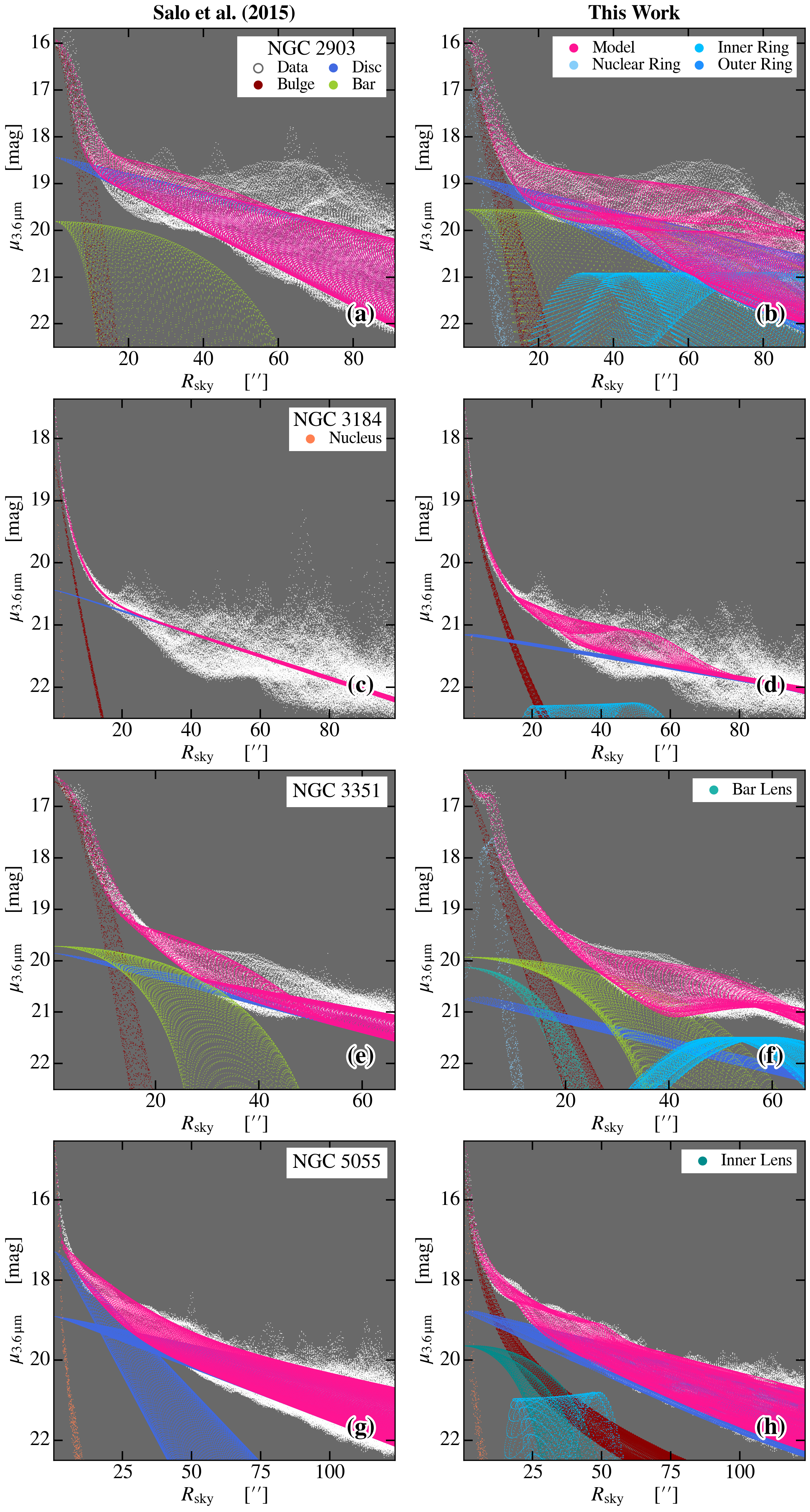}
    \caption{2D profiles constructed from the best-fit models presented by \cite{2015APJS..219....4S} and those constructed in the present work. The surface brightness of each individual components are shown on top of the back-ground data (white) and the best-fit model as a whole (magenta), and as a function of the sky-projected separation $R_{\mathrm{sky}}$. Left columns shows the results from \cite{2015APJS..219....4S}, which is to be compared with those from the present work shown in the right column. (a, b) \ngc{2903}, (c, d) \ngc{3184}, (e, f) \ngc{3351}, (g, h) \ngc{5055}.}
    \label{fig:compare_decomp_0}
\end{figure}

\begin{figure}
    \centering
    \includegraphics[width=0.95\columnwidth]{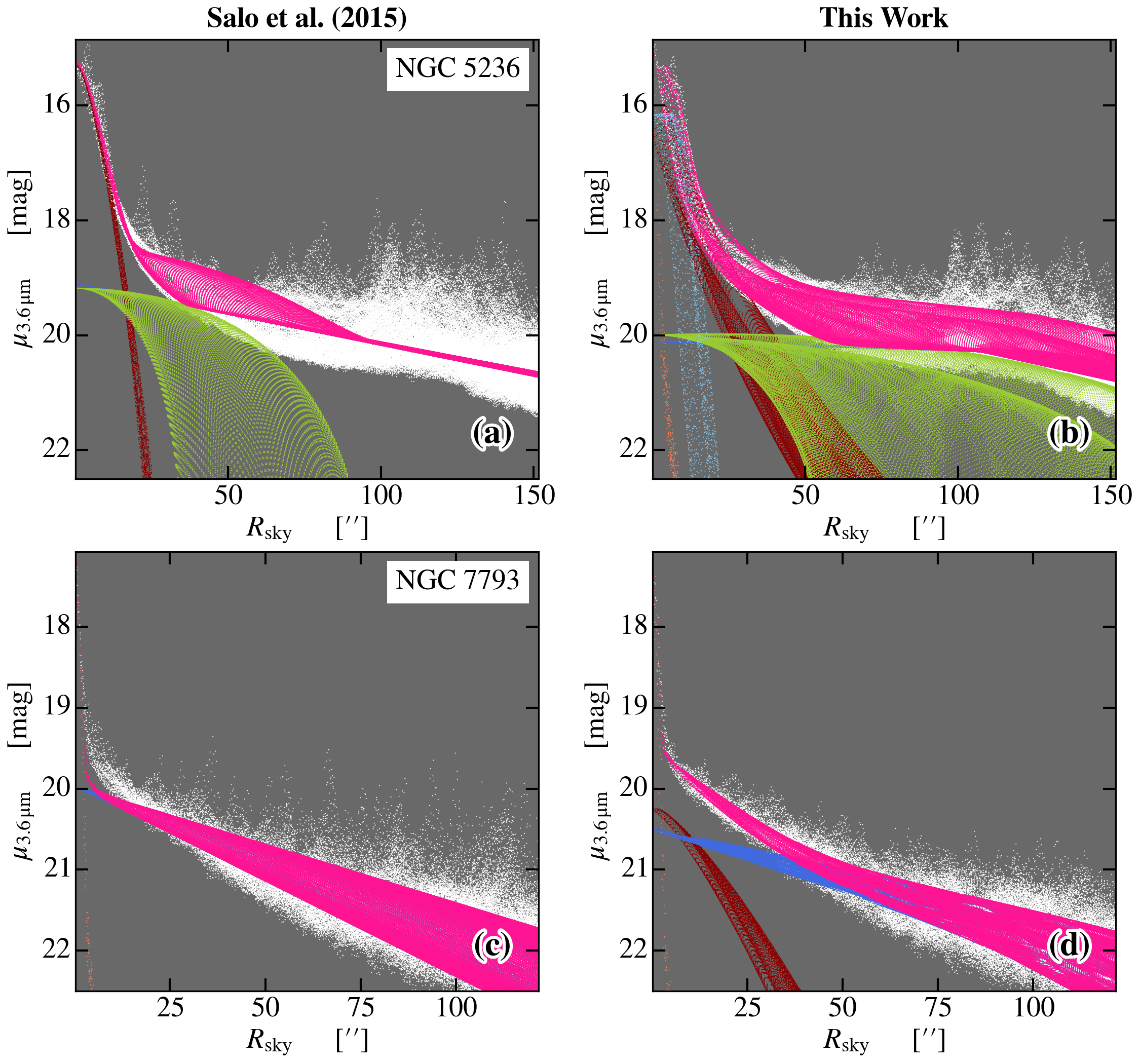}
    \caption{Figure~\ref{fig:compare_decomp_0} continued. (a, b) \ngc{5236}, (c, d) \ngc{7793}.}
    \label{fig:compare_decomp_1}
\end{figure}

Figure~\ref{fig:compare_bulge} suggests that including secondary morphological features (nuclei, rings, lenses, and disk breaks, etc.) could occasionally lead to significant differences in the multi-component decomposition when compared to more simplified treatments. Figure~\ref{fig:compare_decomp_0}--\ref{fig:compare_decomp_1} display some of the most extreme cases corresponding to the objects specifically labelled in Figure~\ref{fig:compare_bulge}, for which differences exceeding $\sim \qty{0.3}{\dex}$ are seen in at least one bulge parameter.

In Figure~\ref{fig:compare_decomp_0}--\ref{fig:compare_decomp_1}, 2D profiles corresponding to optimal multi-component decomposition models constructed in the present work are compared to those presented by \cite{2015APJS..219....4S}. In general, it could be seen from Figure~\ref{fig:compare_decomp_0}--\ref{fig:compare_decomp_1} that including components such as nuclear rings (\ngc{2903}, \ngc{3351} and \ngc{5236}) and bar lenses \cite[\ngc{3351},][]{2015A&A...582A..86H, 2015MNRAS.454.3843A} improves the fitting, such that the consequent best-fit models are seen to agree with the data in a more detailed manner. The improvement is seen not only in terms of the surface brightness but also in terms of the corresponding scatter at a given sky-projected separation, especially in the central regions.

Two objects in our sample, i.e., \ngc{5055} [Figure~\ref{fig:compare_decomp_0}(g, h)] and \ngc{7793} [Figure~\ref{fig:compare_decomp_1} (c, d)], are considered by \cite{2015APJS..219....4S} to be bulge-less but are modelled with a bulge component included in the present work. Figure~\ref{fig:compare_decomp_0}--\ref{fig:compare_decomp_1} suggest that the best-fit models with bulge are favored for their better agreement with the data, especially at $r\approx$ \qtyrange[range-phrase=--, range-units=bracket]{10}{20}{\arcsec}.

Combined with the fact that the bulge stretching correction could be a major source of uncertainty (Table~\ref{table:uncertainty}), it seems that an elaborate, detailed multi-component decomposition could be crucial to improve the robustness and accuracy of torque field evaluation.

\subsection{Comparison with \cite{2016AAA...587A.160D}} \label{subsec:comparison}

\begin{figure}
    \centering
    \includegraphics[width=\columnwidth]{"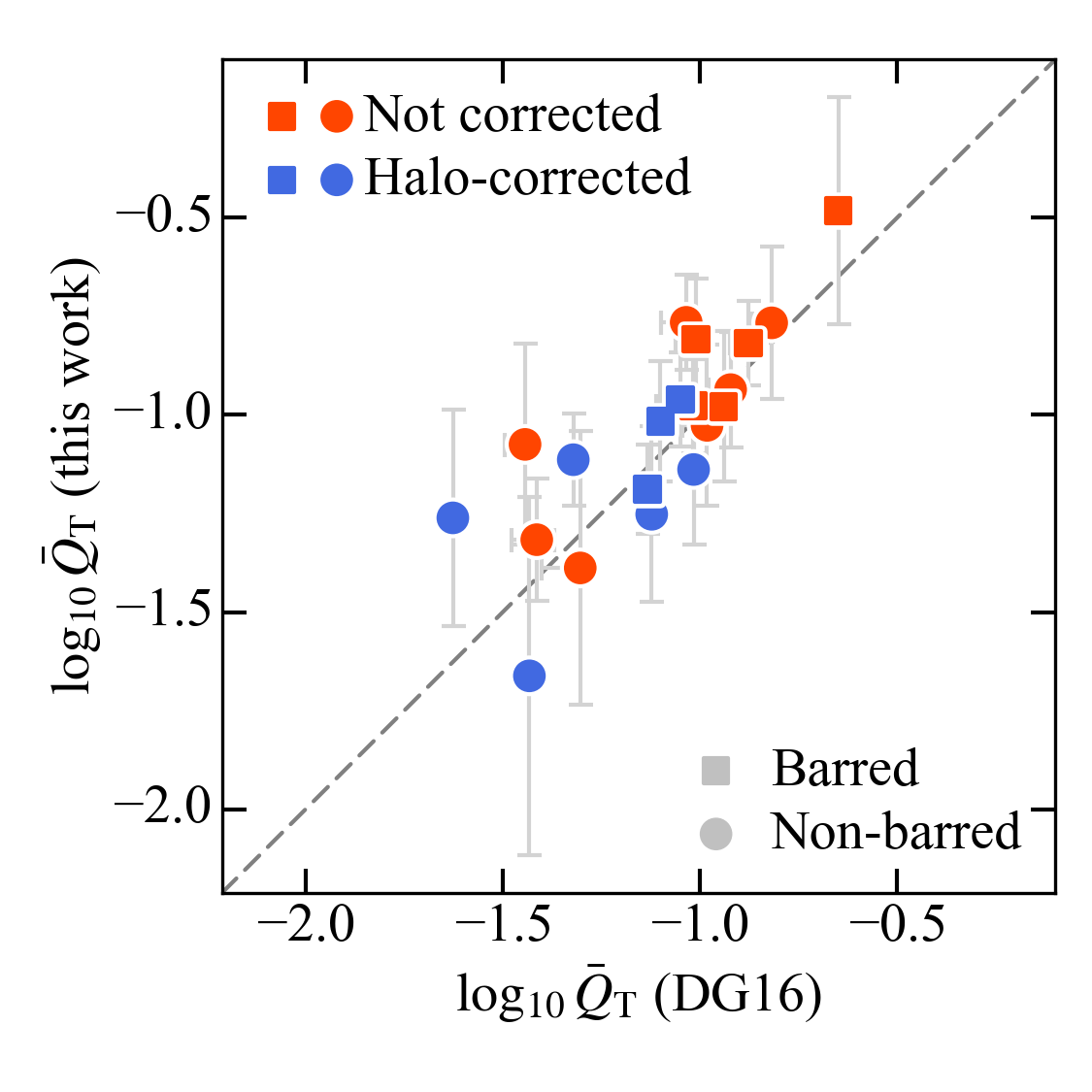"}
    \caption{Comparison between the average torque forces $\bar Q_{\mathrm{T}}$ evaluated in the present work, and that evaluated by \cite{2016AAA...587A.160D}. Symbols are for barred (square) and non-barred (circle) galaxies, and according to whether or not halo correction is applied (red and blue, respectively). The dashed line represents $y=x$. General agreement is seen.}
    \label{fig:comp_measure_dg16}
\end{figure}

For a final check for robustness, we compare our evaluation of the torque force field to that of \cite{2016AAA...587A.160D}, who have evaluated torque force fields for a sample of $\sim \qty{600}{}$ \sfourg galaxies to characterize the stellar bars. Note that \cite{2016AAA...587A.160D} have not applied dark matter halo correction to all their sample galaxies. Also note that \cite{2016AAA...587A.160D} adopted azimuthal maximum instead of median to extract the $Q_{\mathrm{T}}\left( r\right)$ profiles. In these two aspects, we adjust our measurements to comply to their methodology for the purpose of fair comparison.

Figure~\ref{fig:comp_measure_dg16} compares the average torque force $\bar Q_{\mathrm{T}}$ measured in the present work, to that derived from the $Q_{\mathrm{T}}\left( r\right)$ profiles evaluated by \cite{2016AAA...587A.160D}. Differences beyond the estimated uncertainty are only occasionally seen with several data-points, and could be attributed to differences in (\romannumeral1) the considered Fourier harmonics. We consider lower odd-order harmonics in addition to even-order harmonics, which explains the slightly larger $\bar Q_{\mathrm{T}}$ compared to \cite{2016AAA...587A.160D} for \ngc{2903} (specifically for $r\gtrsim \qty{100}{\arcsec}$) and \ngc{3184} ($r\gtrsim \qty{50}{\arcsec}$), since both are multi-armed and a signifiant contribution from odd-order Fourier harmonics could thus be expected \cite[]{2015ApJS..217...32B, 2018ApJ...862...13Y}; (\romannumeral2) the disk orientation. For \ngc{5055}, \cite{2016AAA...587A.160D} have adopted the disk inclination estimated from outer isophotes \cite[]{2015APJS..219....4S}, while we adopt an inclination derived from the \ion{H}{1} kinematic model \cite[]{2008AJ....136.2648D}.

Despite these detailed differences, general agreement within uncertainty is seen between the two measurements shown in Figure~\ref{fig:comp_measure_dg16}, suggesting that our measurements are largely robust.

\section{Partial Correlations Controlling for Neutral Gas Fraction as Covariant} \label{sec:partial_gas_fraction}

\begin{figure*}
    \centering
    \includegraphics[width=\linewidth]{"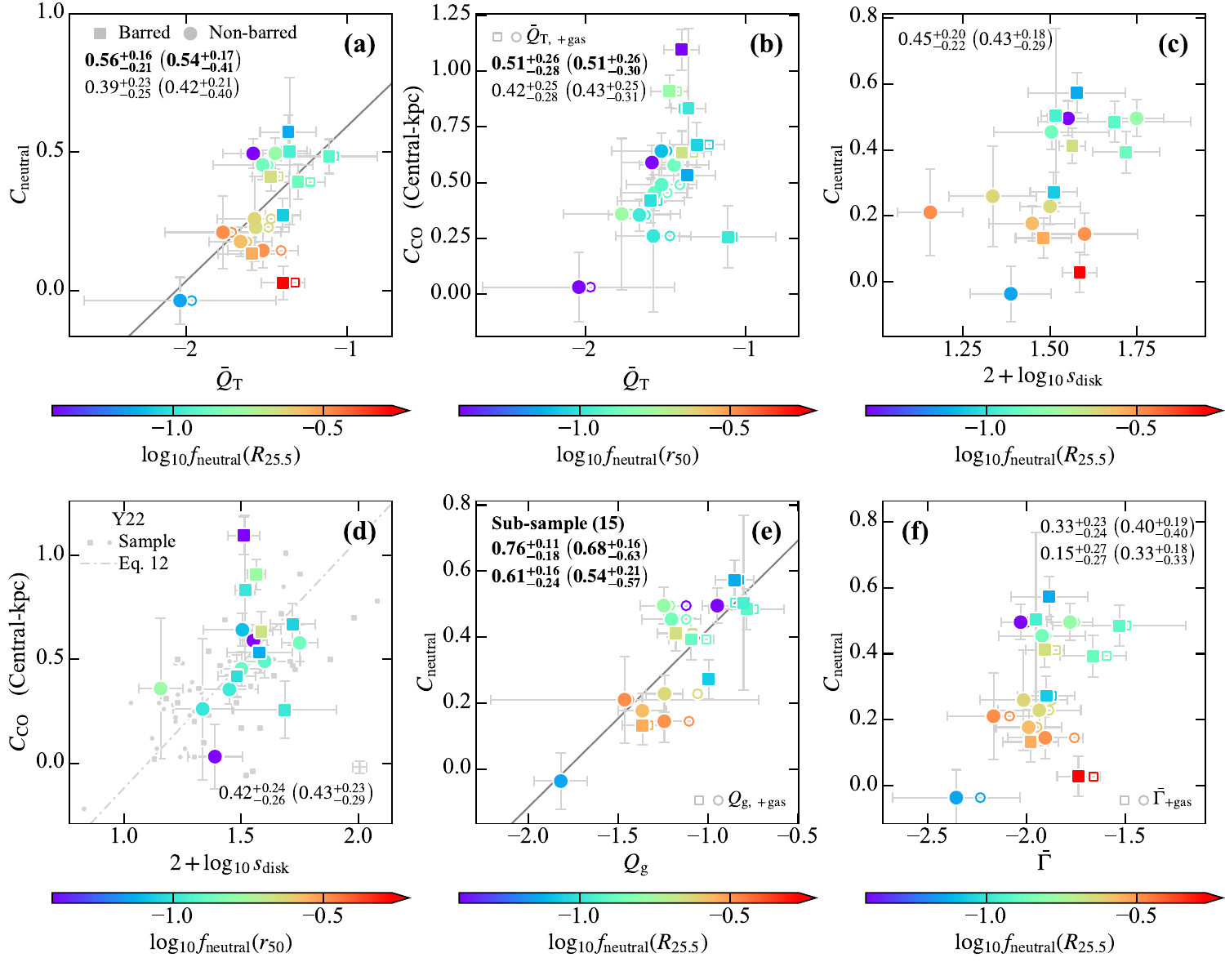"}
    \caption{Correlations discussed in Section~\ref{sec:result}--Section~\ref{sec:discussion}, color-coded instead by the neutral gas fraction $f_{\mathrm{neutral}}\left( R_{25.5}\right)$ (within $R_{25.5}$) or $f_{\mathrm{neutral}}\left( r_{50}\right)$ (within $r_{50}$). Spearman's and partial correlation coefficients (controlling for the neutral gas fraction within the corresponding radial range, bracketed) are quoted at the corner of each panel. In panels (a, b, e, f), correlation coefficients for torque forces evaluated without (first row) and with gas self-gravity considered (second row) are quoted, respectively. Statistically significant ($p<\qty{0.05}{}$) correlations are highlighted in bold typeface.}
    \label{fig:final}
\end{figure*}

Figure~\ref{fig:final} shows alternative versions of Figure~\ref{fig:result:im_c05}--Figure~\ref{fig:result:corrs}, as well Figure~\ref{fig:result:tq_c}, in which the symbols are color-coded by the neutral gas fraction $f_{\mathrm{neutral}}\left( R_{25.5}\right)$ (within $R_{25.5}$) or $f_{\mathrm{neutral}}\left( r_{50}\right)$ (within $r_{50}$). Spearman's and partial correlation coefficients controlling for neutral gas fraction within the corresponding radial range are quoted in Figure~\ref{fig:final}.

\section{Measurements for Individual Galaxies} \label{sec:individual}


\begin{splitdeluxetable*}{ccccccccccccBcccccccccc}
    \tabletypesize{\footnotesize}
    \tablewidth{0pt}
    \tablecaption{Summary of Measurements}
    \label{table:measure}
    \tablehead{
        \colhead{Index}
        & \colhead{Object}
        & \multicolumn{2}{c}{Average Torque Forces}
        & \multicolumn{6}{c}{Maximal Torque Forces}
        & \multicolumn{2}{c}{Average Torques}
        & \multicolumn{3}{c}{Normalized Fourier Amplitudes}
        & \colhead{$s_\mathrm{disk}$}
        & \multicolumn{2}{c}{Gas Concentration Indices}
        & \multicolumn{2}{c}{Neutral Gas Fractions}
        & \colhead{$\bar \Sigma _\mathrm{\star}$}
        & \colhead{Note}
        \\
        \cline{3-4}
        \cline{5-10}
        \cline{11-12}
        \cline{13-15}
        \cline{17-18}
        \cline{19-20}
        \colhead{} &
        \colhead{} &
        \colhead{$\bar Q_{\mathrm{T}}$} &
        \colhead{$\bar Q_{\mathrm{T,\, +gas}}$} &
        \colhead{$Q_{\mathrm{g}}$} &
        \colhead{$r_{Q_\mathrm{g}}$} &
        \colhead{Gas peak} &
        \colhead{$Q_{\mathrm{g,\, +gas}}$} &
        \colhead{$r_{Q_\mathrm{g},\, \mathrm{+gas}}$} &
        \colhead{Gas peak} &
        \colhead{$\bar \Gamma$} &
        \colhead{$\bar \Gamma _{\mathrm{+gas}}$} &
        \colhead{$\max \left\{ A_{2}\right\}$} &
        \colhead{$r_{\max \left\{ A_{2}\right\}}$} &
        \colhead{Gas peak} &
        \colhead{} &
        \colhead{$C_\mathrm{neutral}$} &
        \colhead{$C_\mathrm{CO}$} &
        \colhead{$f_\mathrm{neutral}\left( R_{25.5}\right)$} &
        \colhead{$f_\mathrm{neutral}\left( r_{50}\right)$} &
        \colhead{(\qty{}{\solarmass \per \square \parsec})} &
        \colhead{}
        \\
        \colhead{} &
        \colhead{} &
        \colhead{} &
        \colhead{} &
        \colhead{} &
        \colhead{(\qty{}{\arcsecond})} &
        \colhead{(\qty{}{\arcsecond})} &
        \colhead{} &
        \colhead{(\qty{}{\arcsecond})} &
        \colhead{(\qty{}{\arcsecond})} &
        \colhead{} &
        \colhead{} &
        \colhead{} &
        \colhead{(\qty{}{\arcsecond})} &
        \colhead{(\qty{}{\arcsecond})} &
        \colhead{} &
        \colhead{} &
        \colhead{} &
        \colhead{} &
        \colhead{} &
        \colhead{} &
        \colhead{}
    }
    \colnumbers
    \startdata
    1 & \ngc{628} & $\qtyerr{0.022}{0.010}$ & \qty{0.024}{} & $\qtyerr{0.04}{0.01}$ & $\qtyerr{128}{16}$ & $\qtyerr{77}{16}$ & \qty{0.04}{} & \qty{111}{} & $\qtyerr{77}{16}$ & $\qtyerr{0.010}{0.004}$ & \qty{0.011}{} & $\qtyerr{0.51}{0.29}$ & $\qtyerr{298}{21}$ & $\qtyerr{77}{16}$ & $\qtyerr{0.28}{0.08}$ & $\qtyerr{0.18}{0.05}$ & $\qtyerr{0.36}{0.07}$ & $\qtyerr{0.27}{0.02}$ & $\qtyerr{0.10}{0.01}$ & $\qtyerr{89}{4}$ &  \\
    2 & \ngc{925} & $\qtyerr{0.040}{0.013}$ & \qty{0.047}{} & --- & --- & --- & --- & --- & --- & $\qtyerr{0.018}{0.005}$ & \qty{0.022}{} & $\qtyerr{0.48}{0.05}$ & $\qtyerr{270}{9}$ & --- & $\qtyerr{0.38}{0.04}$ & $\qtyerr{0.03}{0.06}$ & $\qtyerr{0.63}{0.10}$ & $\qtyerr{0.53}{0.05}$ & $\qtyerr{0.22}{0.02}$ & $\qtyerr{20}{1}$ & *, $\ddagger$ \\
    3 & \ngc{2403} & $\qtyerr{0.026}{0.012}$ & \qty{0.028}{} & $\qtyerr{0.04}{0.01}$ & $\qtyerr{168}{41}$ & $\qtyerr{217}{6}$ & \qty{0.05}{} & \qty{168}{} & $\qtyerr{217}{6}$ & $\qtyerr{0.010}{0.003}$ & \qty{0.012}{} & $\qtyerr{0.45}{0.03}$ & $\qtyerr{196}{8}$ & $\qtyerr{217}{6}$ & $\qtyerr{0.30}{0.06}$ & $\qtyerr{0.13}{0.06}$ & $\qtyerr{0.42}{0.10}$ & $\qtyerr{0.29}{0.03}$ & $\qtyerr{0.10}{0.01}$ & $\qtyerr{60}{3}$ &  \\
    4 & \ngc{2903} & $\qtyerr{0.050}{0.019}$ & \qty{0.059}{} & $\qtyerr{0.08}{0.02}$ & $\qtyerr{156}{46}$ & --- & \qty{0.10}{} & \qty{55}{} & --- & $\qtyerr{0.022}{0.008}$ & \qty{0.025}{} & $\qtyerr{0.78}{0.14}$ & $\qtyerr{178}{16}$ & --- & $\qtyerr{0.52}{0.12}$ & $\qtyerr{0.39}{0.06}$ & $\qtyerr{0.67}{0.10}$ & $\qtyerr{0.13}{0.01}$ & $\qtyerr{0.09}{0.01}$ & $\qtyerr{239}{23}$ & $\ddagger$ \\
    5 & \ngc{2976} & $\qtyerr{0.077}{0.053}$ & \qty{0.083}{} & $\qtyerr{0.16}{0.08}$ & $\qtyerr{65}{11}$ & $\qtyerr{77}{4}$ & \qty{0.18}{} & \qty{58}{} & $\qtyerr{77}{4}$ & $\qtyerr{0.029}{0.023}$ & \qty{0.032}{} & $\qtyerr{0.75}{0.38}$ & $\qtyerr{96}{19}$ & $\qtyerr{77}{4}$ & $\qtyerr{0.49}{0.25}$ & $\qtyerr{0.49}{0.06}$ & $\qtyerr{0.26}{0.14}$ & $\qtyerr{0.12}{0.01}$ & $\qtyerr{0.10}{0.02}$ & $\qtyerr{72}{11}$ &  \\
    6 & \ngc{3031} & $\qtyerr{0.009}{0.013}$ & \qty{0.011}{} & $\qtyerr{0.02}{0.01}$ & $\qtyerr{280}{268}$ & $\qtyerr{605}{210}$ & \qty{0.02}{} & \qty{293}{} & $\qtyerr{605}{210}$ & $\qtyerr{0.004}{0.003}$ & \qty{0.006}{} & $\qtyerr{0.47}{0.10}$ & $\qtyerr{615}{30}$ & $\qtyerr{605}{210}$ & $\qtyerr{0.24}{0.07}$ & $\qtyerr{-0.04}{0.08}$ & $\qtyerr{0.03}{0.15}$ & $\qtyerr{0.07}{0.01}$ & $\qtyerr{0.00}{0.00}$ & $\qtyerr{418}{30}$ &  \\
    7 & \ngc{3184} & $\qtyerr{0.027}{0.013}$ & \qty{0.032}{} & $\qtyerr{0.06}{0.01}$ & $\qtyerr{82}{25}$ & $\qtyerr{98}{9}$ & \qty{0.09}{} & \qty{75}{} & $\qtyerr{98}{9}$ & $\qtyerr{0.012}{0.006}$ & \qty{0.013}{} & $\qtyerr{0.51}{0.15}$ & $\qtyerr{213}{37}$ & $\qtyerr{98}{9}$ & $\qtyerr{0.32}{0.06}$ & $\qtyerr{0.23}{0.06}$ & $\qtyerr{0.45}{0.08}$ & $\qtyerr{0.23}{0.02}$ & $\qtyerr{0.13}{0.01}$ & $\qtyerr{61}{4}$ &  \\
    8 & \ngc{3351} & $\qtyerr{0.040}{0.010}$ & \qty{0.041}{} & $\qtyerr{0.10}{0.01}$ & $\qtyerr{47}{4}$ & $\qtyerr{70}{0}$ & \qty{0.10}{} & \qty{51}{} & $\qtyerr{70}{0}$ & $\qtyerr{0.013}{0.004}$ & \qty{0.013}{} & $\qtyerr{0.52}{0.02}$ & $\qtyerr{39}{4}$ & $\qtyerr{70}{0}$ & $\qtyerr{0.32}{0.05}$ & $\qtyerr{0.27}{0.06}$ & $\qtyerr{1.10}{0.09}$ & $\qtyerr{0.09}{0.01}$ & $\qtyerr{0.04}{0.00}$ & $\qtyerr{217}{13}$ &  \\
    9 & \ngc{3621} & $\qtyerr{0.027}{0.014}$ & \qty{0.034}{} & --- & --- & --- & --- & --- & --- & $\qtyerr{0.010}{0.005}$ & \qty{0.013}{} & $\qtyerr{0.25}{0.14}$ & $\qtyerr{226}{9}$ & --- & $\qtyerr{0.22}{0.06}$ & $\qtyerr{0.26}{0.15}$ & $\qtyerr{0.26}{0.34}$ & $\qtyerr{0.24}{0.05}$ & $\qtyerr{0.10}{0.04}$ & $\qtyerr{139}{7}$ & *, $\dagger$, $\ddagger$ \\
    10 & \ngc{3627} & $\qtyerr{0.043}{0.017}$ & \qty{0.042}{} & $\qtyerr{0.14}{0.03}$ & $\qtyerr{50}{12}$ & $\qtyerr{70}{13}$ & \qty{0.15}{} & \qty{48}{} & $\qtyerr{70}{13}$ & $\qtyerr{0.013}{0.006}$ & \qty{0.013}{} & $\qtyerr{0.58}{0.12}$ & $\qtyerr{141}{16}$ & $\qtyerr{70}{13}$ & $\qtyerr{0.38}{0.12}$ & $\qtyerr{0.57}{0.06}$ & $\qtyerr{0.53}{0.10}$ & $\qtyerr{0.07}{0.01}$ & $\qtyerr{0.07}{0.01}$ & $\qtyerr{244}{34}$ &  \\
    11 & \ngc{4736} & $\qtyerr{0.026}{0.011}$ & \qty{0.027}{} & $\qtyerr{0.11}{0.02}$ & $\qtyerr{8}{2}$ & $\qtyerr{42}{0}$ & \qty{0.08}{} & \qty{15}{} & $\qtyerr{42}{0}$ & $\qtyerr{0.009}{0.002}$ & \qty{0.010}{} & $\qtyerr{0.59}{0.08}$ & $\qtyerr{201}{4}$ & $\qtyerr{315}{6}$ & $\qtyerr{0.36}{0.05}$ & $\qtyerr{0.50}{0.05}$ & $\qtyerr{0.59}{0.08}$ & $\qtyerr{0.04}{0.00}$ & $\qtyerr{0.02}{0.00}$ & $\qtyerr{2021}{57}$ &  \\
    12 & \ngc{5055} & $\qtyerr{0.030}{0.021}$ & \qty{0.032}{} & $\qtyerr{0.06}{0.02}$ & $\qtyerr{79}{11}$ & --- & \qty{0.08}{} & \qty{82}{} & --- & $\qtyerr{0.012}{0.006}$ & \qty{0.013}{} & $\qtyerr{0.58}{0.08}$ & $\qtyerr{493}{3}$ & --- & $\qtyerr{0.32}{0.12}$ & $\qtyerr{0.45}{0.05}$ & $\qtyerr{0.64}{0.08}$ & $\qtyerr{0.14}{0.01}$ & $\qtyerr{0.08}{0.01}$ & $\qtyerr{195}{13}$ & $\ddagger$ \\
    13 & \ngc{5194} & $\qtyerr{0.036}{0.018}$ & \qty{0.037}{} & $\qtyerr{0.06}{0.02}$ & $\qtyerr{79}{16}$ & $\qtyerr{21}{9}$ & \qty{0.06}{} & \qty{70}{} & $\qtyerr{21}{9}$ & $\qtyerr{0.017}{0.004}$ & \qty{0.018}{} & $\qtyerr{0.94}{0.08}$ & $\qtyerr{317}{8}$ & $\qtyerr{21}{9}$ & $\qtyerr{0.56}{0.10}$ & $\qtyerr{0.50}{0.06}$ & $\qtyerr{0.58}{0.09}$ & $\qtyerr{0.16}{0.01}$ & $\qtyerr{0.13}{0.02}$ & $\qtyerr{235}{21}$ &  \\
    14 & \ngc{5236} & $\qtyerr{0.044}{0.011}$ & \qty{0.041}{} & $\qtyerr{0.16}{0.01}$ & $\qtyerr{99}{10}$ & $\qtyerr{112}{0}$ & \qty{0.14}{} & \qty{78}{} & $\qtyerr{112}{0}$ & $\qtyerr{0.011}{0.002}$ & \qty{0.011}{} & $\qtyerr{0.47}{0.01}$ & $\qtyerr{115}{12}$ & $\qtyerr{112}{0}$ & $\qtyerr{0.33}{0.03}$ & $\qtyerr{0.50}{0.26}$ & $\qtyerr{0.83}{0.36}$ & $\qtyerr{0.11}{0.04}$ & $\qtyerr{0.10}{0.05}$ & $\qtyerr{339}{6}$ & $\dagger$ \\
    15 & \ngc{5457} & $\qtyerr{0.030}{0.015}$ & \qty{0.039}{} & $\qtyerr{0.06}{0.02}$ & $\qtyerr{254}{34}$ & $\qtyerr{266}{31}$ & \qty{0.08}{} & \qty{127}{} & $\qtyerr{126}{20}$ & $\qtyerr{0.012}{0.005}$ & \qty{0.017}{} & $\qtyerr{0.68}{0.35}$ & $\qtyerr{511}{32}$ & $\qtyerr{490}{47}$ & $\qtyerr{0.40}{0.14}$ & $\qtyerr{0.15}{0.06}$ & $\qtyerr{0.49}{0.08}$ & $\qtyerr{0.32}{0.04}$ & $\qtyerr{0.12}{0.02}$ & $\qtyerr{59}{6}$ &  \\
    16 & \ngc{6946} & $\qtyerr{0.034}{0.009}$ & \qty{0.037}{} & $\qtyerr{0.07}{0.02}$ & $\qtyerr{51}{8}$ & --- & \qty{0.08}{} & \qty{150}{} & --- & $\qtyerr{0.012}{0.003}$ & \qty{0.014}{} & $\qtyerr{0.70}{0.03}$ & $\qtyerr{464}{5}$ & --- & $\qtyerr{0.37}{0.03}$ & $\qtyerr{0.41}{0.05}$ & $\qtyerr{0.91}{0.07}$ & $\qtyerr{0.22}{0.02}$ & $\qtyerr{0.17}{0.02}$ & $\qtyerr{195}{6}$ & $\ddagger$ \\
    17 & \ngc{7793} & $\qtyerr{0.017}{0.014}$ & \qty{0.019}{} & $\qtyerr{0.03}{0.06}$ & $\qtyerr{49}{22}$ & --- & \qty{0.04}{} & \qty{144}{} & --- & $\qtyerr{0.007}{0.004}$ & \qty{0.008}{} & $\qtyerr{0.21}{0.02}$ & $\qtyerr{152}{10}$ & --- & $\qtyerr{0.14}{0.03}$ & $\qtyerr{0.21}{0.13}$ & $\qtyerr{0.36}{0.34}$ & $\qtyerr{0.33}{0.06}$ & $\qtyerr{0.17}{0.05}$ & $\qtyerr{60}{1}$ & $\dagger$, $\ddagger$ \\
   \enddata
    \tablecomments{Summary of measurements. (1) Index. (2) Object designation. (3) Average torque forces $\bar Q_{\mathrm{T}}$. (4) Average torque force $\bar Q_{\mathrm{T,\, +gas}}$, with the neutral gas self-gravity combined in evaluating the torque forces. (5) Maximal torque forces $Q_{\mathrm{g}}$. (6) The radius $r_{Q_{\mathrm{g}}}$ corresponding to $Q_{\mathrm{g}}$. (7) Radius of \qty{}{\kilo \parsec}-scale neutral gas overdensity proximate to $r_{Q_{\mathrm{g}}}$. (8--10) Same as (5--7), with the neutral gas self-gravity combined in evaluating the torque forces. (11) Average torque $\bar \Gamma$. (12) Average torque $\bar \Gamma _{\mathrm{+gas}}$, with the neutral gas self-gravity combined in evaluating the torque. (13) Maximal $m=2$ Fourier amplitude of the stellar mass surface density $\max \left\{ A_2\right\}$. (14) Radius $r_{\max \left\{ A_2\right\}}$ corresponding to $\max \left\{ A_2\right\}$. (15) Radius of \qty{}{\kilo \parsec}-scale neutral gas overdensity proximate to $r_{\max \left\{ A_2\right\}}$. (16) Average stellar mass surface density Fourier amplitudes $s_{\mathrm{disk}}$. (17) Inner neutral gas concentration parameter $C_\mathrm{neutral}$. (18) Central molecular gas concentration index $C_\mathrm{CO}$. (19) Neutral gas fraction $f_\mathrm{neutral}\left( R_{25.5}\right)$ within $R_{25.5}$. (20) Neutral gas fraction $f_\mathrm{neutral}\left( r_{50}\right)$ within $r_{50}$. (21) Average stellar mass surface density $\bar \Sigma _\mathrm{\star}$. (22) Note. \\
        *. Maximal torque forces are not determined given the overall monotonically decreasing $Q_{\mathrm{T}}\left( r\right)$ [$Q_{\mathrm{T,\, +gas}}\left( r\right)$] as a function of radius. \\
        $\dagger$. Molecular gas distribution is inferred from the star formation rate distribution by inverting the sub-\qty{}{\kilo \parsec} Schmidt law \cite[]{2008AJ....136.2846B}. \\
        $\ddagger$. \qty{}{\kilo \parsec}-scale neutral gas overdensity cannot determined from the generally smooth neutral gas radial profile.
    }
\end{splitdeluxetable*}


\begin{figure*}
    \centering
    \includegraphics[width=\linewidth]{"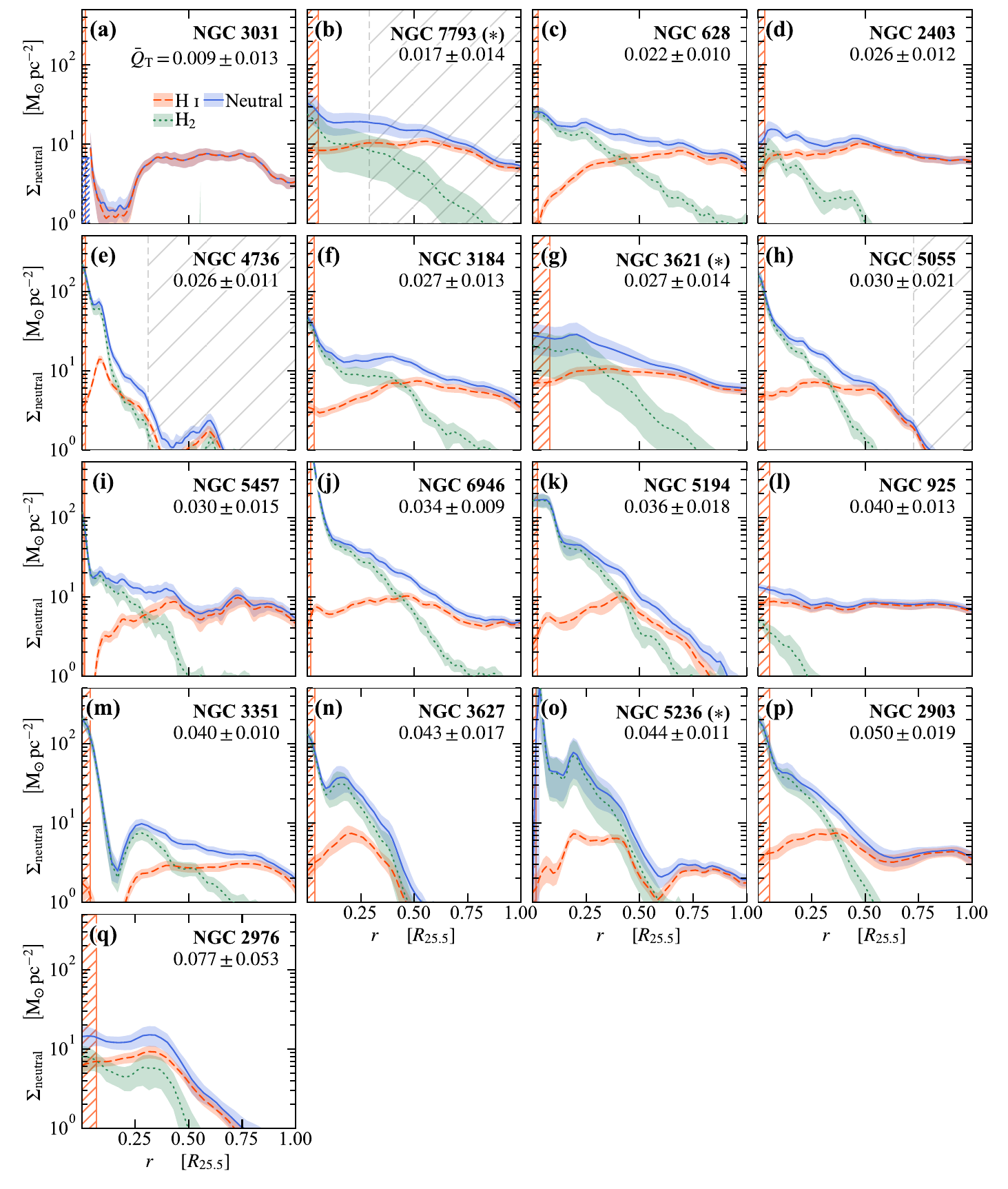"}
    \caption{Neutral gas surface density profiles (blue solid curve) for the sample galaxies, as well as the corresponding \ion{H}{1} (orange dashed curve) and molecular gas (green dotted curve) profiles. For galaxies whose designations are shown with asterisks, molecular gas distributions are derived by inverting the sub-\qty{}{\kilo \parsec}-scale Schmidt law \cite[]{2008AJ....136.2846B}. Hatched regions correspond to \ion{H}{1} disk warps. In this figure, panels are sorted by the increasing average torque forces $\bar Q_{\mathrm{T}}$, labeled at the top right-hand corner. Moving from top to bottom, the average torque forces $\bar Q_{\mathrm{T}}$ increase, while the neutral gas distribution appears to be more concentrated at $\sim \qty{0.5}{}R_{25.5}$-scale.}
    \label{fig:neutral_gas_profile}
\end{figure*}

\begin{figure*}
    \centering
    \includegraphics[width=0.8\linewidth]{"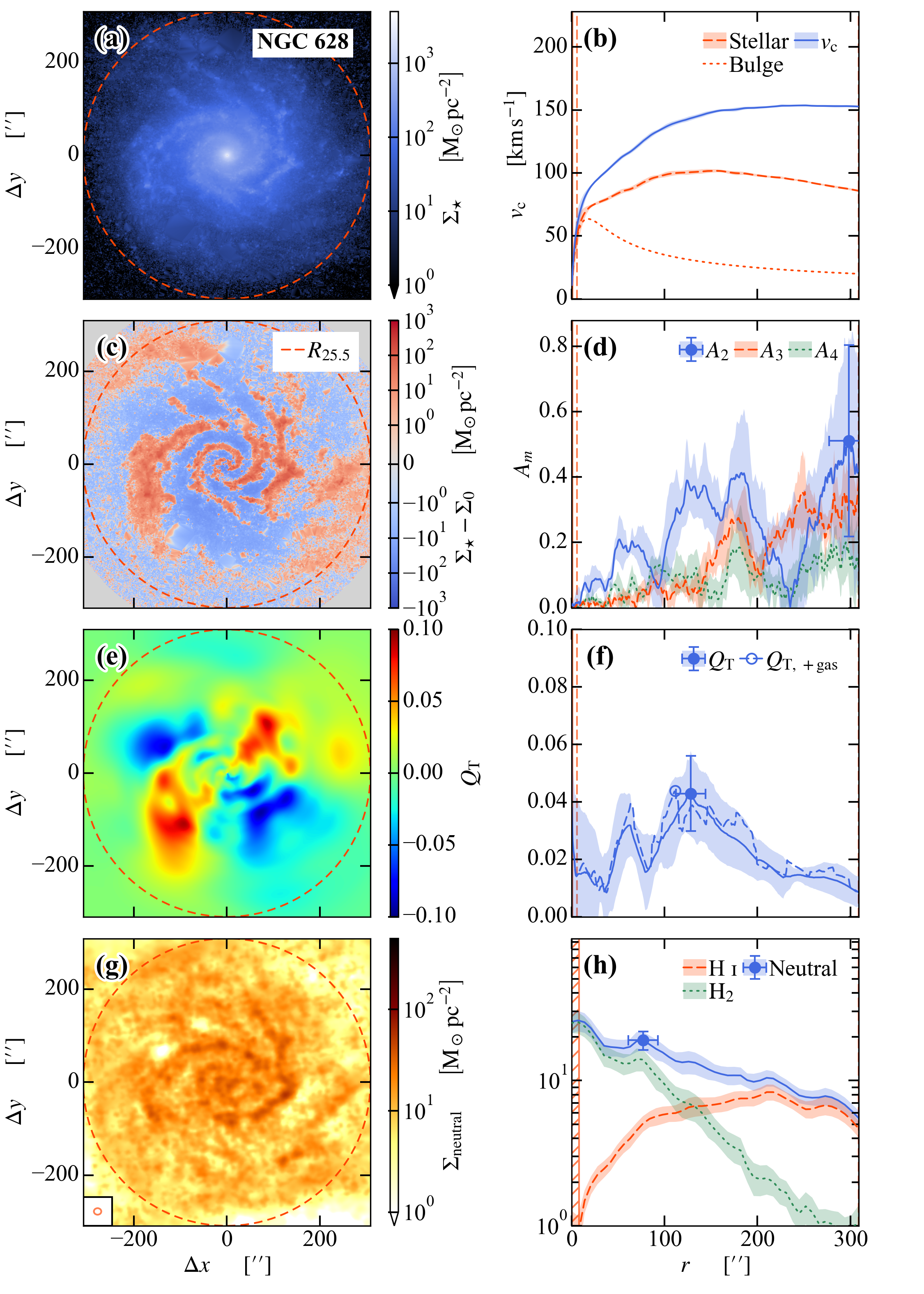"}
    \caption{Summary of measurements for \ngc{628}. The left column shows (a) the de-projected stellar mass surface density $\Sigma _\star$ with (c) the non-axisymmetric structures highlighted, (e) the evaluated torque force field $Q_{\mathrm{T}}$, and (g) the neutral gas distribution. Red circles indicate $R_{25.5}$. Right column shows (b) the corresponding rotational velocity, (d) Fourier amplitudes, (f) torque forces, and (h) neutral gas surface density as functions of radius. In panel (b), rotational velocity solely due to the stellar mass (red dashed curve) and more specifically due to the stellar bulge (red dotted curve) are shown along the estimated rotation curve $v_{\mathrm{c}}$ (blue solid curve). In panel (d), Fourier amplitudes $A_2$ (blue), $A_3$ (red) and $A_4$ (green) are shown, with the $\max \left\{ A_2\right\}$ highlighted with the blue symbol. In panel (f), torque force due to stars $Q_{\mathrm{T}}$ (blue solid curve), and that evaluated with the neutral gas self-gravity combined $Q_{\mathrm{T,\, +gas}}$ (blue dashed curve) are shown, with the maximal value $Q_{\mathrm{g}}$ ($Q_{\mathrm{g,\, +gas}}$) highlighted by the filled (empty) symbol, respectively. In panel (h), neutral gas mass surface density $\Sigma _\mathrm{neutral}$ (blue solid curve), \ion{H}{1} mass surface density $\Sigma _\text{\ion{H}{1}}$ (red dashed curve) and molecular gas mass surface density $\Sigma _\mathrm{H_2}$ (green dotted curve) are shown, with \qty{}{\kilo \parsec}-scale neutral gas over-densities highlighted by the blue symbols. \\
    \textbf{The complete figure set (17 images) is available in the online journal.}}
    \label{fig:summary_ngc0628}
\end{figure*}

We show the neutral gas profiles for each individual sample galaxies in Figure~\ref{fig:neutral_gas_profile}. For convenience of comparison, panels in Figure~\ref{fig:neutral_gas_profile} are ordered according to the average torque force $\bar Q_{\mathrm{T}}$. We summarize the measured physical quantities in Table~\ref{table:measure}, and show the measured torque field distribution and neutral gas distribution in Figure~\ref{fig:summary_ngc0628}, for each of the individual galaxies.

\figsetstart
\figsetnum{14}
\figsettitle{Summary of measurements for individual galaxies.}
\label{figset:summary}

\figsetend


\bibliography{2023a}{}
\bibliographystyle{aasjournal}



\end{document}